# Brief Large Language Model–Based Chatbot Conversations and Parental Intentions for HPV Vaccination for Children: A Randomized Clinical Trial

Neil K. R. Sehgal, ME[1,2], Sunny Rai, PhD[1,2], Manuel Tonneau, MSc[3,4,5], Anish K. Agarwal MD,MPH, MS[2,6,7], Joseph Cappella, PhD[2,8], Melanie Kornides, ScD, RN, FNP-BC[2,9], Lyle Ungar, PhD[1], Alison Buttenheim, PhD, MBA[2,9,10], Sharath Chandra Guntuku, PhD[1,2]

[1]Computer and Information Science Department, University of Pennsylvania, Philadelphia, PA, USA
[2]Leonard Davis Institute of Health Economics, University of Pennsylvania, Philadelphia, PA, USA
[3]Oxford Internet Institute, University of Oxford, Oxford, United Kingdom
[4]Development Impact Department, World Bank, Washington, DC, USA
[5]Department of Computer Science, New York University, New York, NY, USA
[6]Penn Medicine Center for Health Care Transformation and Innovation, University of Pennsylvania, Philadelphia, PA, USA
[7]Department of Emergency Medicine, University of Pennsylvania Perelman School of Medicine, Philadelphia, PA, USA
[8]Annenberg School of Communication, University of Pennsylvania, Philadelphia, PA, USA
[9]School of Nursing, University of Pennsylvania, Philadelphia, PA, USA
[10]Center for Health Incentives and Behavioral Economics, University of Pennsylvania, Philadelphia, PA, USA.

**Corresponding Author:** Neil K. R. Sehgal, ME

**Email:** nsehgal@upenn.edu

**Competing Interest Statement:** The authors have no disclosures to report

**Trial Registration:** ClinicalTrials.gov Identifier: NCT07132125

**Key Points**

**Question**
Do brief conversations with a large language model (LLM) chatbot increase parental intention to vaccinate children against human papillomavirus (HPV) more than standard public health materials?

**Findings**
In this preregistered randomized clinical trial of 1297 parents of unvaccinated children, brief LLM chatbot interactions significantly increased HPV vaccination intent compared with no intervention but did not outperform public health materials, and their effects did not persist by 45 days.

**Meaning**
Brief LLM chatbot conversations may increase short-term vaccination intentions but offer limited advantage over existing public health messaging for promoting HPV vaccination.

**ABSTRACT**

**Importance**
Health systems are increasingly considering large language model (LLM)–based chatbots for vaccine communication, but evidence they improve durable, behaviorally relevant outcomes beyond existing health materials is limited.

**Objective**
To test whether brief, multi-turn interactions with an LLM chatbot increase parental intention to vaccinate children against human papillomavirus (HPV), compared with no intervention and government public health materials, and whether any effects persist.

**Design**
Preregistered randomized controlled trial conducted March–April 2025, with follow-up at 15 and 45 days.

**Setting**
Online study conducted in the United States, Canada, and the United Kingdom.

**Participants**
Parents 18 years or older with at least one HPV vaccine–eligible child (ages 11–17 years in the US and Canada; 12–17 years in the UK) who had not received HPV vaccination or whose vaccination status was unknown.

**Interventions**
Participants were randomized to (1) no-message control, (2) government public health materials matched to country (minimum 3-minute exposure), or (3) a 3-minute interaction with an LLM chatbot (GPT-4o) prompted to encourage HPV vaccination using either a default response style or a shorter conversational style.

**Main Outcomes and Measures**
The primary outcome was self-reported likelihood of vaccinating the child against HPV within 12 months (0–100 scale) measured immediately post-intervention. Prespecified follow-ups included vaccination intent and self-reported vaccination at 15 and 45 days.

**Results**
A total of 1297 participants were randomized; mean participant age was 42.8 years, and 72.1% were women. Compared with no intervention, public health materials increased immediate vaccination intent (Cohen's d, 0.53; 95% CI, 0.36–0.70), as did the default chatbot (d, 0.48; 95% CI, 0.30–0.65) and conversational chatbot (d, 0.33; 95% CI, 0.17–0.49). At 45 days, neither chatbot increased intent relative to controls, whereas public health materials maintained modest effects. No intervention increased self-reported vaccination uptake.

**Conclusions and Relevance**
In this randomized controlled trial, we found well-designed public health messaging may match or exceed the impact of short chatbot conversations for HPV vaccine promotion.

**Trial Registration**
ClinicalTrials.gov Identifier: NCT07132125

## Introduction

Vaccination against human papillomavirus (HPV) prevents approximately 90% of HPV-associated cancers, including cervical and oropharyngeal cancers.[1] Despite its high effectiveness, the vaccine remains significantly underutilized globally.[2] U.S. adolescent vaccination rates have stagnated at around 60%, far behind other adolescent immunizations such as for tetanus and meningococcal disease.[3] In addition to structural barriers such as cost and access, parental hesitancy is one of the key barriers towards improving coverage.[4,5]

In response, health systems and public agencies are increasingly exploring artificial intelligence–based tools to support vaccine communication.[6,7] Large language model (LLM) based chatbots have garnered significant attention for their potential to shape beliefs, attitudes, and decision-making by generating highly persuasive messaging. Early studies across domains such as political persuasion, misinformation correction, and health communication find that LLMs can change beliefs and attitudes, oftentimes more effectively than humans and at scale.[8-16] However, many existing studies rely on weak or minimal control conditions and focuses on immediate attitudinal outcomes rather than behaviorally relevant endpoints or more distal outcomes.

Evidence for chatbot-based vaccine promotion is similarly mixed. Some studies report short-term improvements in health intentions when chatbots are compared with no intervention or brief informational controls, while others show limited or no effects on downstream behaviors.[17,18] Recent randomized trials in pneumococcal vaccination and HIV self-testing in China have reported improvements in behavioral outcomes such as vaccine uptake or testing completion.[19,20] However, both trials evaluated multi-component or service-delivery chatbots — hybrid or rule-based systems rather than LLMs — embedded within structured health service workflows. These systems differ substantially from brief, one-time LLM interactions, and it is unclear how they perform relative to high-quality public health materials that reflect current practice.

HPV vaccination provides a useful test case for evaluating the real-world value of LLM-based communication. Unlike COVID-19 vaccination, HPV vaccination is less politically polarized and involves a finite dosing schedule rather than recurring boosters, simplifying the interpretation of parental intentions.[21]

In this preregistered randomized controlled trial, we tested whether a brief, three-minute, multi-turn interaction with an LLM-driven chatbot increases parental intentions to vaccinate their child against HPV, relative to both no intervention and government public health materials. We tested two chatbot variants, a default LLM output style and a more conversational style. To assess durability, we measured vaccination intentions immediately after the intervention and again at 15 and 45 days, and we examined self-reported vaccination and related behavioral outcomes. This design allowed us to evaluate whether LLM chatbots offer meaningful advantages over existing public health resources for HPV vaccine promotion.

## Methods

### Study Design and Participants

We conducted a preregistered randomized clinical trial to evaluate whether brief interactions with a large language model chatbot could increase parental intention to vaccinate their child against human papillomavirus. Parents residing in the United States, Canada, or the United Kingdom were recruited through two online platforms known for providing high-quality respondent samples (Prolific and CloudResearch Connect) between March and April 2025.[22] Eligible participants were aged 18 years or older and had at least one child eligible for HPV vaccination (ages 11–17 years in the United States and Canada; ages 12–17 years in the United Kingdom) who had not received any doses of the HPV vaccine or who were unsure on their child's vaccination status. Participants were compensated at a rate equivalent to US $18 per hour.

The study consisted of a baseline survey (Survey 1), followed by an intervention and immediate post-intervention assessment approximately one week later (Survey 2), with follow-up surveys administered 15 and 45 days after the intervention. This approach aimed to reduce demand characteristics and expectancy bias.

The baseline survey collected data on demographics, child vaccination history, trust in public health institutions and AI, general vaccine hesitancy using the PACV-5 scale,[23] and self-reported likelihood of vaccinating their child for HPV, influenza, and COVID-19 within the next 12 months (0–100 scale). To identify a consistent index child, participants were first asked to provide the first initial of their youngest HPV vaccine-eligible child. All subsequent child-specific questions referred to this index child using the initial provided. Reasons for HPV non-vaccination were adapted from the U.S. National Immunization Survey–Teen survey (NIS-Teen); participants rated all reasons endorsed by more than 1% of respondents in the NIS-Teen 2018 survey on a 0–100 scale where 0 was marked "Not important at all" and 100 was marked "Extremely important", in addition to an open-ended "other" response option.[24] We separated the NIS-Teen category "more information/new vaccine" into two items ("needing more information" and "the vaccine being too new") and did not include a direct analogue of the broader "lack of knowledge" category, which was considered conceptually overlapping with these informational items. These items were analyzed as individual indicators rather than as a composite scale.

All participants provided informed consent electronically via Qualtrics prior to participation. The study was approved by the University of Pennsylvania Institutional Review Board, preregistered at AsPredicted (aspredicted.org/m7tt-y2y4.pdf), and retrospectively registered at ClinicalTrials.gov (NCT07132125) (see Appendix for both). The preregistration specified the primary outcome (post-intervention HPV vaccination intent), follow-up outcomes (intent at 15 and 45 days and vaccination uptake), key secondary outcomes, exclusion criteria (attention check failure and implausibly short completion time), and the primary regression-based analytic approach. The ClinicalTrials.gov record was created after study completion and reflects the same primary and secondary outcomes and study design; no primary or secondary outcomes were added or modified, and no additional exclusion criteria were introduced. As described below, the primary analysis deviated from the preregistered exclusion criteria by retaining all randomized participants in an intention-to-treat framework and handling missing data via multiple imputation. The overall study flow and timing of surveys, interventions, and follow-ups are shown in Figure S1. This study is reported in accordance with the CONSORT reporting guidelines for randomized controlled trials (see Supplement for checklist).

**Randomization and Interventions**

Participants who completed Survey 1 and met the eligibility criteria were randomly assigned to one of four conditions: (1) no-message control, (2) government public health materials (strong control), (3) LLM chatbot with default response style, or (4) LLM chatbot with a conversational response style.

In the strong control condition, participants read government-issued HPV vaccination materials matched to their country of residence (Centers for Disease Control and Prevention, National Health Service, or Public Health Agency of Canada).[25-28] Materials ranged from 589 to 680 words and were displayed for a minimum of three minutes before participants were able to proceed. Readability of the materials corresponded to grade 9.3 (United States), 8.3 (Canada), and 6.1 (United Kingdom) based on Flesch-Kincaid Grade Levels. Full stimuli are provided in the Supplement (Appendix).

In the chatbot conditions, participants engaged in a three-minute minimum, multi-turn conversation with an LLM-based chatbot implemented using OpenAI's GPT-4o (gpt-4o-2024-11-20) model. Chatbots were instructed to encourage parents to ensure their child receives the HPV vaccine and were provided each participant's top-rated reason for HPV non-vaccination from the baseline survey. The default chatbot used the model's standard output style (i.e. often using paragraphs and bullet points), whereas the conversational chatbot was instructed to respond in a conversational tone using no more than three sentences per message. Prior work finds that informal, conversational cues can boost a chatbot's social presence, the feeling of interacting with a real person, which is associated with greater engagement, trust, and perceived interaction quality, and may increase receptivity to persuasive health messages.[29,30] No fine-tuning, rule-based scripting, or additional training data were used. A three-minute session length was chosen to mirror realistic, practical constraints for public-health use cases. Full chatbot instruction prompts and example exchanges are provided in Tables S1–S2. Examples of the chatbot interface are provided in Figures S2-S3.

**Outcomes**

The primary pre-registered outcome was self-reported likelihood of vaccinating the participant's child against HPV within the next 12 months, measured immediately after the intervention on a 0–100 scale (0 = extremely unlikely; 100 = extremely likely).

Pre-registered follow-up outcomes included HPV vaccination intent at 15 and 45 days post-intervention, as well as HPV vaccination uptake. Parents who reported that their child had received the HPV vaccine at follow-up were routed past the vaccination intent item by survey skip logic, and assigned a follow-up intent score of 100, reflecting behavioral resolution of the outcome. Other pre-registered secondary outcomes included general parental vaccine hesitancy (PACV-5), intentions to vaccinate the child against influenza and COVID-19 in the next 12 months, willingness to donate a portion of study compensation to an immunization-related charity, and click-through to a pharmacy vaccine scheduling link (United States participants only). Participants in the intervention arms also rated perceived empathy, usefulness, and relevance, as well as message effectiveness (adapted from an existing public health message effectiveness scale)[24] of the material or conversation.

**Statistical Analysis**

The primary analysis followed an intention-to-treat framework and included all 1297 randomized participants. The primary estimand was the effect of assigned study arm on immediate post-intervention HPV vaccination intent.

Continuous intent outcomes were analyzed using analysis-of-covariance style linear regression with HC2 robust standard errors, adjusting for baseline HPV intent. Standardized effect sizes (Cohen's d) were calculated by dividing adjusted arm contrasts by the pooled residual standard deviation. Binary outcomes were analyzed using logistic regression with baseline HPV intent adjustment, and we report adjusted odds ratios with marginal risk differences.

Missing postbaseline outcomes were handled using multiple imputation with 50 imputed data sets and predictive mean matching for bounded continuous outcomes. The imputation model included assigned arm, baseline HPV vaccination intent, baseline COVID and influenza vaccination intent, baseline PACV-5 hesitancy score, country, recruitment platform, age, gender, race/ethnicity, education, baseline child HPV vaccination uncertainty, postrandomization survey-completion flags, child HPV vaccination status at each follow-up wave, and observed post-randomization outcomes (immediate, 15-day, and 45-day intent; process evaluation ratings).
Because intent was measured on a bounded 0–100 scale, we examined arm-specific distributions, residual-vs-fitted plots, Q-Q plots, and influence diagnostics, and we fit fractional-logit sensitivity models after rescaling intent to the unit interval. We also evaluated differential attrition by arm using chi-square tests and logistic regression, compared baseline characteristics of completers vs non-completers using standardized mean differences, and fit inverse-probability-weighted sensitivity models for the primary outcome.

Regression diagnostics were conducted to assess model assumptions, including Breusch-Pagan and White tests for heteroskedasticity, and D'Agostino-Pearson and Shapiro-Wilk tests for residual normality. Breusch-Pagan and White tests indicated heteroskedasticity (both $p<0.05$), confirming the appropriateness of HC2-robust standard errors. As an additional sensitivity analysis for the bounded [0,100] scale, fractional logistic regression was estimated for the primary outcome contrast (default chatbot vs. no-message control); inverse probability weighting (IPW) was also estimated to cross-validate the primary OLS approach.

As an exploratory analysis, we evaluated whether treatment effects on immediate postintervention HPV vaccination intent differed across baseline participant characteristics. Using the randomized ITT cohort and the same multiple-imputation framework as the primary analysis, we fit separate linear models for each baseline moderator that included study arm, baseline HPV vaccination intent, the moderator main effect, and arm-by-moderator interaction terms. For categorical moderators, interaction effects were evaluated with omnibus Wald tests across all interaction terms for that moderator; p values were adjusted for multiple testing across moderators using the Benjamini-Hochberg false discovery rate procedure.

Our pre-registered analysis plan specified exclusion of participants who failed embedded attention checks or whose survey completion times were implausibly short. In the primary analysis reported here, we instead retain all 1,297 randomized participants in an intention-to-treat framework, handling missing outcome data from non-completers via multiple imputation (described above). This approach constitutes a minor deviation from the pre-registered exclusion criteria for attention checks and completion time; these flags are retained as descriptive covariates rather than exclusion criteria. Beyond the pre-registered criteria, we also tracked several additional post-hoc data quality indicators — reCAPTCHA scores, child initial mismatches across survey waves, technical issues, and non-entry into Survey 2 — which informed a complete-case sensitivity analysis (Table S5) but were not applied as exclusions in the primary ITT analysis. For these participants, we also conducted exploratory analyses examining the conversational and persuasive strategies used by the chatbots and their associations with post-intervention vaccination intent (details in Appendix).

All tests were two-sided, with $p < .05$ considered statistically significant. Additional methodological details, including screening criteria and exploratory content analyses, are provided in the Appendix.

**Results**

**Participant Characteristics and Study Flow**

A total of 1297 participants were randomized: 324 to no message, 325 to public health materials, 324 to the default chatbot, and 324 to the conversational chatbot (Figure S4). Participants in the randomized cohort had a mean age of 42.8 years, 72.1% identified as women, and 63.6% resided in the United States;

baseline characteristics were well balanced across conditions (Table 1). Mean baseline likelihood of vaccinating the index child against HPV was 27.4 (SD, 34.23) on a 0–100 scale, reflecting substantial baseline hesitancy.

Immediate primary-outcome data were observed for 1096 participants (290 no message, 283 public health, 262 default chatbot, and 261 conversational chatbot). Fifteen-day intent data were observed for 952 participants (249, 241, 225, and 237, respectively), and 45-day intent data were observed for 893 participants (230, 226, 208, and 229, respectively).

**Intervention Fidelity**

All participants assigned to the public health materials condition who provided an observed response (285 of 325) reported reading the materials; 98% reported reading them carefully. Participants in the chatbot conditions engaged actively, sending a mean of 63.2 (SD, 43.4) words during the interaction, while the default and conversational chatbots generated 820.6 (329.1) and 372.6 (162.4) words respectively (Table S3). Interactions with the conversational chatbot involved more frequent but shorter exchanges than those with the default chatbot.

**Primary Outcome: Immediate HPV Vaccination Intent**

Across all conditions, HPV vaccination intent increased by a mean of 9.2 points (SD, 25.0) from baseline to the immediate post-intervention assessment.

Relative to the no intervention control, all active interventions increased immediate vaccination intent (Figure 1; Table 2, Table S3). Government public health materials produced the largest increase (Cohen's d, 0.53; 95% CI, 0.36–0.70). The default chatbot also increased intent (d, 0.48; 95% CI, 0.30–0.65), while the conversational chatbot yielded a smaller effect (d, 0.33; 95% CI, 0.17–0.49). Results were qualitatively similar when restricting to a complete-case sample (n=930) applying both pre-registered quality screens (attention check failure, implausibly short completion time) and additional post-hoc data integrity criteria (Table S5). No significant moderation by participant demographics or baseline beliefs was detected (Table S6).

Regression diagnostics indicated mild positive skewness in OLS residuals (skewness=0.62) and evidence of heteroskedasticity (Breusch-Pagan LM=12.3, p=0.02), supporting the use of HC2-robust standard errors. Substantive conclusions were robust across all three contrasts in sensitivity analyses using fractional logistic regression, which explicitly models the bounded [0,100] scale, and inverse probability weighting (Table S7).

**Attrition**

Attrition at the immediate post-intervention timepoint differed significantly across arms ($\chi^2$ = 14.97; P = .002). Both chatbot arms showed significantly lower odds of providing primary outcome data relative to no message (default chatbot: OR, 0.50; P = .002; conversational chatbot: OR, 0.49; P = .002), while the public health arm did not differ significantly from no message (OR, 0.79; P = .34). The primary driver of differential attrition in chatbot arms was participants who entered the intervention survey but did not complete the chatbot interaction and subsequent intent item: 39 participants in the default chatbot arm and 33 in the conversational chatbot arm entered the intervention survey but did not provide primary outcome data, compared with 11 in the no-message arm and 9 in the public health arm.

Non-completers had a slightly higher mean PACV-5 hesitancy score at baseline (3.57 [SD, 0.98]) relative to completers (3.41 [SD, 1.04]; SMD, −0.16; P = .04), suggesting marginally greater baseline hesitancy among those lost to the immediate follow-up. Baseline HPV intent (27.57 [SD, 34.63] vs 26.55 [SD, 32.02]; P = .68) and age (42.81 [SD, 6.97] vs 43.01 [SD, 6.75]; P = .71) did not differ significantly. Overall, observed baseline differences were modest in magnitude, consistent with, though not definitive evidence for, a missing-at-random assumption (Table S8).

**Follow-up Outcomes at 15 and 45 Days**

At 15 days post-intervention, both the conversational chatbot and public health materials increased HPV vaccination intent relative to the no message control (Figure 1, Table 2). In contrast, the default chatbot did not differ from the no message control. No intervention increased self-reported HPV vaccination at 15 days (Table 3).

By 45 days, only the public health materials increased HPV vaccination intent relative to the no message control (Cohen's d, 0.17; 95% CI, 0.06-0.28) (Figure 1, Table 2). No intervention increased self-reported HPV vaccination uptake at 45 days (Table 3).

Effects at day 15 and day 45 were robust to a sensitivity analysis excluding participants whose child had been vaccinated by follow-up (n=63 at day 15; n=82 at day 45), who were assigned an intent score of 100 in the primary analysis (Table S9).

**Secondary Outcomes**

None of the interventions produced spillover effects on influenza or COVID-19 vaccination intentions, general parental vaccine hesitancy (PACV-5), willingness to donate to an immunization-related charity, or click rates on a vaccine scheduling link (United States participants only) (Tables 2-3).

Across 13 prespecified reasons for non-vaccination, four were decreased by at least one chatbot or the public health reading material (Figure S5). All active interventions significantly decreased the mean score across the 13 reasons relative to the no message control.

Participants rated the default chatbot as more empathetic than the public health materials. The conversational chatbot was rated as less effective than the public health materials on an existing message effectiveness scale. Participants spent more time engaging with the default chatbot than with the public health materials (mean difference, 69 seconds), despite the absence of larger or more durable effects (Table S10).

Exploratory analyses examined the conversational and persuasive strategies used by the chatbots and their associations with post-intervention vaccination intent among the per-protocol sample. Details are provided in the Appendix, including principal component and regularized regression analyses, comparisons with public health materials, and evaluations of adherence to motivational interviewing principles (Tables S11 – S16; Figure S6-S7).

**Discussion**

In this preregistered randomized clinical trial, brief, three-minute conversations with an LLM-based chatbot increased parental intentions to vaccinate their child against HPV immediately after exposure but did not produce durable effects and did not outperform existing government public health materials. In contrast, public health materials yielded modest but durable increases in vaccination intent through 45 days. None of the interventions increased self-reported HPV vaccination or related behavioral outcomes.

The absence of vaccination uptake effects across all arms reflects a combination of factors. HPV vaccination requires scheduling, may involve a pediatrician visit, and 45 days may be too short for intent to convert to action; even among parents whose intentions shifted, structural barriers such as access and competing priorities remained unaddressed. Lastly, recruitment fell modestly short of our pre-specified sample size, which may have reduced statistical power to detect effects at follow-up timepoints and for rare binary outcomes such as self-reported vaccination uptake. This underpowering may have limited our ability to detect true differences at 45 days, and results at that timepoint should be interpreted with appropriate caution.

Still, our findings suggest that, at least in their current form and with brief exposure, LLM chatbots may not provide meaningful advantages over established public health communication resources for promoting HPV vaccination. While chatbots offer interactive and personalized exchanges, these features did not translate into greater or more persistent changes in vaccination intentions when compared with existing well-designed, evidence-based materials.

Two recent HPV-specific chatbot trials provide a useful contrast. In a U.S. randomized trial, brief interactions with a GPT-4–based chatbot increased parental HPV vaccination intentions relative to a CDC brochure, although parents spent substantially more time with the chatbot than with the control materials and outcomes were assessed within the same survey session.[18] In a separate school-based cluster trial in China, parents given two weeks of on-demand access to an LLM-powered chatbot were substantially more likely to schedule or obtain HPV vaccination for their children compared with no intervention.[31] In contrast, our study required equivalent minimum exposure across conditions, used existing public health materials as a strong comparator, introduced a delay between baseline and intervention assessment, and evaluated durability over 45 days. These design differences may help explain why we observed smaller and less persistent effects of chatbot-based communication.

The lack of durable effects is consistent with a broader literature showing that brief informational or persuasive interventions often struggle to produce lasting changes in vaccine-related decisions, particularly when structural barriers and contextual factors remain unaddressed.[32] Although HPV vaccination is less politically polarized than COVID-19 vaccination, parental decisions still involve logistical considerations, competing priorities, and trust in medical institutions.[4] These factors may not be readily shifted through short conversational encounters alone. This pattern is consistent with recent work in cancer screening, where an LLM-based chatbot increased willingness to complete stool-based colorectal cancer screening but did not affect intentions to undergo colonoscopy, a more invasive and logistically demanding procedure.[17]

Exploratory analyses of chatbot conversations, reported in the Supplement, offer tentative insights into how different messaging approaches may relate to short-term changes in intention. Strategies emphasizing external validation and collective benefits were associated with higher post-intervention intent, whereas more personalized empathy-focused exchanges were associated with lower intent. These findings should be interpreted cautiously, as conversational content was shaped by participant responses and cannot be disentangled from underlying motivation.

**Limitations**

This study has several limitations. Vaccination intentions and uptake were self-reported and may be subject to social desirability or recall bias. Participants were recruited from online platforms and may not generalize to all parents with low HPV vaccination intent, particularly those with limited digital access or more entrenched hesitancy. The sample was also predominantly White and female, which limits conclusions about whether chatbot- or materials-based communication would perform differently across demographic groups. Parents encountered in clinic-based settings may differ meaningfully from online panel respondents in their baseline motivation, institutional trust, and receptivity to health messaging. Furthermore, the sample skewed toward moderate-to-high AI familiarity, and parents with lower digital literacy may engage with chatbot-based tools differently in ways that could attenuate or amplify the effects observed here. In addition, participants were required to engage with the chatbots or read the public health materials for at least three minutes, which may exceed feasible real-world engagements. However, parents naturally encountering chatbots "in the wild" might be more motivated to engage longer, possibly producing greater benefits than observed here. In fact, participants chose to spend 70 seconds more on average with the default chatbot than with the public health material. It bears emphasizing that shifts on a continuous 0–100 intention scale do not necessarily translate into meaningful behavior change; the absence of vaccination uptake effects across all arms is the more behaviorally consequential finding, and future work should prioritize behavioral endpoints over attitudinal ones.

Despite these limitations, our findings have practical implications for clinicians, health systems, and public health agencies considering the deployment of AI-based communication tools. High-quality, evidence-based public health materials performed at least as well as brief chatbot interactions and demonstrated greater durability. Chatbots may still serve as a supplementary resource, particularly for addressing questions or reinforcing existing guidance, but should not blindly be used in place of established communication strategies.

**Conclusions**

In this preregistered randomized clinical trial of parents with low HPV vaccination intent, brief interactions with an LLM-based chatbot increased intentions immediately after exposure but did not produce durable effects or outperform existing public health materials.

Future research should explore optimal strategies for integrating personalized AI interactions alongside structural interventions, ultimately aiming for a multifaceted approach to address deep-rooted vaccine hesitancy effectively. Rigorous comparisons against realistic control conditions and assessments of durability will be essential as AI tools are increasingly incorporated into public health practice.

**Acknowledgments**

This project was funded in part by the University of Pennsylvania's Penn Medicine Communication Research Institute, Penn Global Research and Engagement Fund, and National Institutes of Health, NIH-NIMHD:R01MD018340 awarded to Dr. Guntuku, NIH-NIMH:R01MH132401 awarded to Dr. Buttenheim, and NIH-NCI:R37CA259210 awarded to Dr. Kornides. The funders had no role in the design and conduct of the study; collection, management, analysis, and interpretation of the data; preparation, review, or approval of the manuscript; and decision to submit the manuscript for publication. Neil Sehgal had full access to all the data in the study and takes responsibility for the integrity of the data and the accuracy of the data analysis.

**Data Availability**
The full, anonymized dataset used in this study is available at https://github.com/sehgal-neil/HPV_LLM_Persuasion

**References**

1. CDC, Reasons to Get Vaccinated. Human Papillomavirus (HPV) (2024). Available at: https://www.cdc.gov/hpv/vaccines/reasons-to-get.html [Accessed 14 April 2025].

2. Han J, Zhang L, Chen Y, et al. Global HPV vaccination programs and coverage rates: a systematic review. EClinicalMedicine. 2025;84:103290. Published 2025 Jun 9. doi:10.1016/j.eclinm.2025.103290

3. C. Pingali, National Vaccination Coverage Among Adolescents Aged 13–17 Years — National Immunization Survey-Teen, United States, 2023. MMWR Morb Mortal Wkly Rep 73 (2024).

4. Cooper S, Schmidt BM, Jama NA, et al. Factors that influence caregivers' and adolescents' views and practices regarding human papillomavirus (HPV) vaccination for adolescents: a qualitative evidence synthesis. Cochrane Database Syst Rev. 2025;4(4):CD013430. Published 2025 Apr 15. doi:10.1002/14651858.CD013430.pub2

5. Lin Z, Chen S, Su L, et al. Influences of HPV disease perceptions, vaccine accessibility, and information exposure on social media on HPV vaccination uptake among 11,678 mothers with daughters aged 9-17 years in China: a cross-sectional study. BMC Med. 2024;22(1):328. Published 2024 Aug 13. doi:10.1186/s12916-024-03538-1

6. Addie YO. Vaccines and motherhood: are AI-generated health messages working in Kenya and Nigeria? The Conversation. Published October 6, 2025. https://doi.org/10.64628/AAJ.akjf7gmka

7. Nicolaus T. Public health communicators finding promise in AI tools. The Nation's Health. 2025;55(9):1-19.

8. G. Czarnek, et al., Addressing climate change skepticism and inaction using human-AI dialogues. [Preprint] (2025). Available at: https://osf.io/mqcwj_v1 [Accessed 14 April 2025].

9. T. H. Costello, G. Pennycook, D. G. Rand, Durably reducing conspiracy beliefs through dialogues with AI. Science 385, eadq1814 (2024).

10. K. Hackenburg, H. Margetts, Evaluating the persuasive influence of political microtargeting with large language models. Proceedings of the National Academy of Sciences 121, e2403116121 (2024).

11. J. A. Goldstein, J. Chao, S. Grossman, A. Stamos, M. Tomz, How persuasive is AI-generated propaganda? PNAS Nexus 3, pgae034 (2024).

12. T. H. Costello, G. Pennycook, D. Rand, Just the facts: How dialogues with AI reduce conspiracy beliefs. [Preprint] (2025). Available at: https://osf.io/h7n8u_v1 [Accessed 14 April 2025].

13. F. Salvi, M. H. Ribeiro, R. Gallotti, R. West, On the Conversational Persuasiveness of Large Language Models: A Randomized Controlled Trial. [Preprint] (2024). Available at: http://arxiv.org/abs/2403.14380 [Accessed 14 April 2025].

14. K. Hackenburg, et al., Scaling language model size yields diminishing returns for single-message political persuasion. Proceedings of the National Academy of Sciences 122, e2413443122 (2025).

15. E. Karinshak, S. X. Liu, J. S. Park, J. T. Hancock, Working With AI to Persuade: Examining a Large Language Model's Ability to Generate Pro-Vaccination Messages. Proc. ACM Hum.-Comput. Interact. 7, 116:1-116:29 (2023).

16. Hackenburg K, Tappin BM, Hewitt L, Saunders E, Black S, Lin H, Fist C, Margetts H, Rand DG, Summerfield C. The levers of political persuasion with conversational artificial intelligence. Science. 2025 Dec 4;390(6777):eaea3884.

17. Sehgal NK, Tonneau M, Tan A, Mehta SJ, Buttenheim A, Ungar L, Agarwal AK, Guntuku SC. Effect of Static vs. Conversational AI-Generated Messages on Colorectal Cancer Screening Intent: a Randomized Controlled Trial. arXiv preprint arXiv:2507.08211. 2025 Jul 10.

18. H. G. Xu, et al., Personalized Dialogues with AI Effectively Address Parents' Concerns about HPV Vaccination. [Preprint] (2025). Available at: https://osf.io/gv52j_v1 [Accessed 14 April 2025].

19. Chen S, Yu F, Fang Y, et al. Chatbot-Delivered Real-Time Support to Improve HIV Self-Testing Rates: A Randomized Clinical Trial. JAMA Netw Open. 2025;8(11):e2544821. doi:10.1001/jamanetworkopen.2025.44821

20. Wang Z, Chen S, Poon J, et al. A Hybrid Chatbot to Promote Pneumococcal Vaccination Among Older Adults: A Randomized Clinical Trial. JAMA Netw Open. 2025;8(10):e2535813. doi:10.1001/jamanetworkopen.2025.35813

21. T. A. Santibanez, C. L. Black, T. Zhou, A. Srivastav, J. A. Singleton, Parental hesitancy about COVID-19, influenza, HPV, and other childhood vaccines. Vaccine 42, 126139 (2024).

22. M. N. Stagnaro, et al., Representativeness versus Response Quality: Assessing Nine Opt-In Online Survey Samples. [Preprint] (2024). Available at: https://osf.io/h9j2d_v1 [Accessed 14 April 2025].

23. D. Opel, Identifying, understanding and talking with vaccine-hesitant parents. University of Washington School of Medicine (2014). Available at: https://www.fondation-merieux.org/wp-content/uploads/2017/03/from-package-to-protection-how-do-we-close-global-coverage-gaps-to-optimize-the-impact-of-vaccination-2014-douglas-opel.pdf [Accessed 14 April 2025].

24. K. Sonawane, et al., Trends in Human Papillomavirus Vaccine Safety Concerns and Adverse Event Reporting in the United States. JAMA Network Open 4, e2124502 (2021).

25. CDC, Reasons to Get Vaccinated. Human Papillomavirus (HPV) (2024). Available at: https://www.cdc.gov/hpv/vaccines/reasons-to-get.html [Accessed 14 April 2025].

26. CDC, Impact of the HPV Vaccine. Human Papillomavirus (HPV) (2024). Available at: https://www.cdc.gov/hpv/vaccination-impact/index.html [Accessed 14 April 2025].

27. NHS, HPV vaccine. nhs.uk (2024). Available at: https://www.nhs.uk/vaccinations/hpv-vaccine/ [Accessed 14 April 2025].

28. P. H. A. of Canada, Human papillomavirus (HPV): Spread, prevention and risks. (2025). Available at: https://www.canada.ca/en/public-health/services/diseases/human-papillomavirus-hpv/spread-prevention-risks.html [Accessed 14 April 2025].

29. Araujo T. Living up to the chatbot hype: The influence of anthropomorphic design cues and communicative agency framing on conversational agent and company perceptions. Computers in human behavior. 2018 Aug 1;85:183-9.

30. Liebrecht C, Sander L, Van Hooijdonk C. Too informal? How a chatbot's communication style affects brand attitude and quality of interaction. InInternational Workshop on Chatbot Research and Design 2020 Nov 23 (pp. 16-31). Cham: Springer International Publishing.

31. Z. Hou, et al., A vaccine chatbot intervention for parents to improve HPV vaccination uptake among middle school girls: a cluster randomized trial. Nat Med 1–8 (2025). https://doi.org/10.1038/s41591-025-03618-6.

32. S. Liu, M. R. Durantini, C. Calabrese, F. Sanchez, D. Albarracin, A systematic review and meta-analysis of strategies to promote vaccination uptake. Nat Hum Behav 8, 1689–1705 (2024).

**Figures and Tables**

**A)**

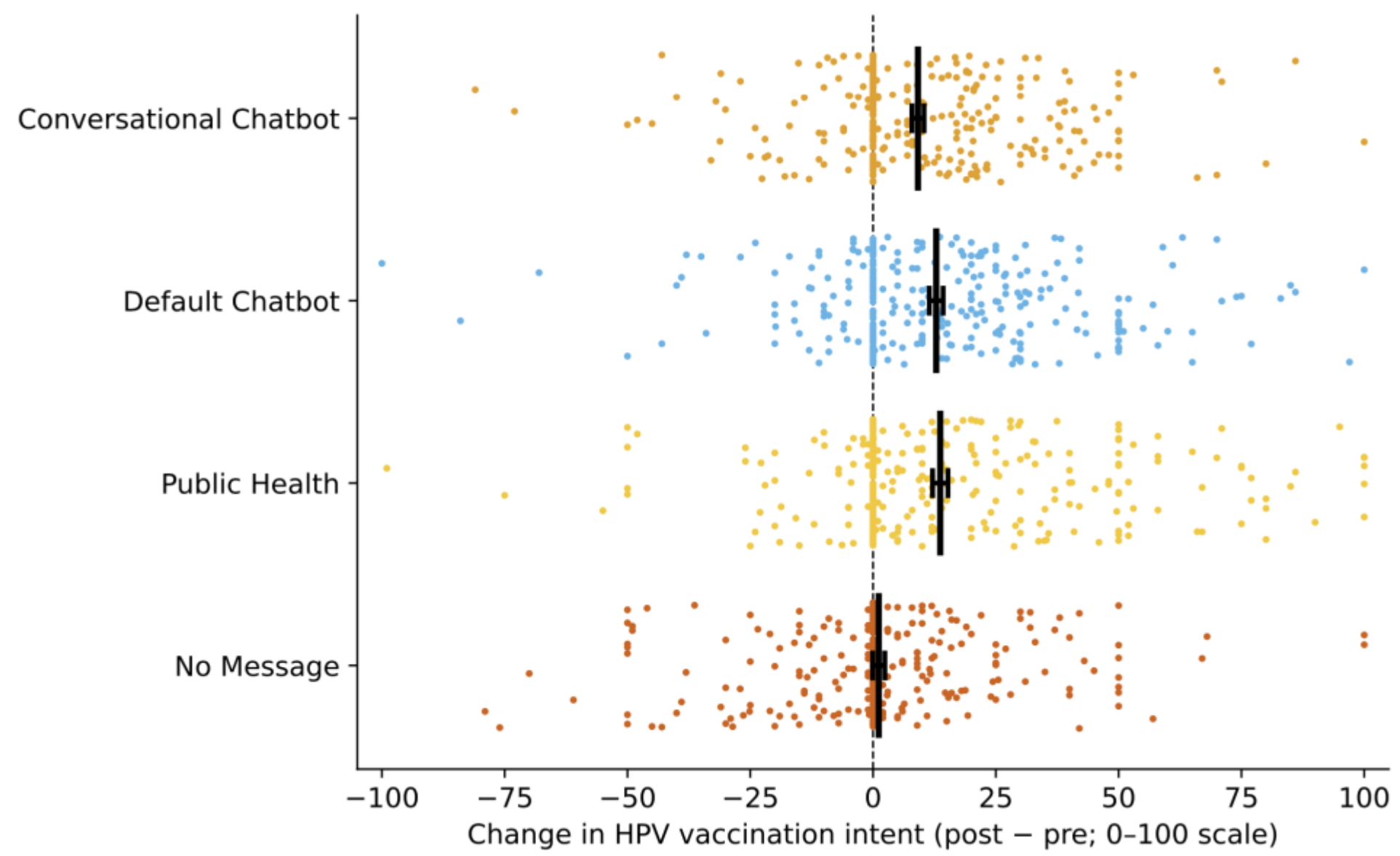

**B)**

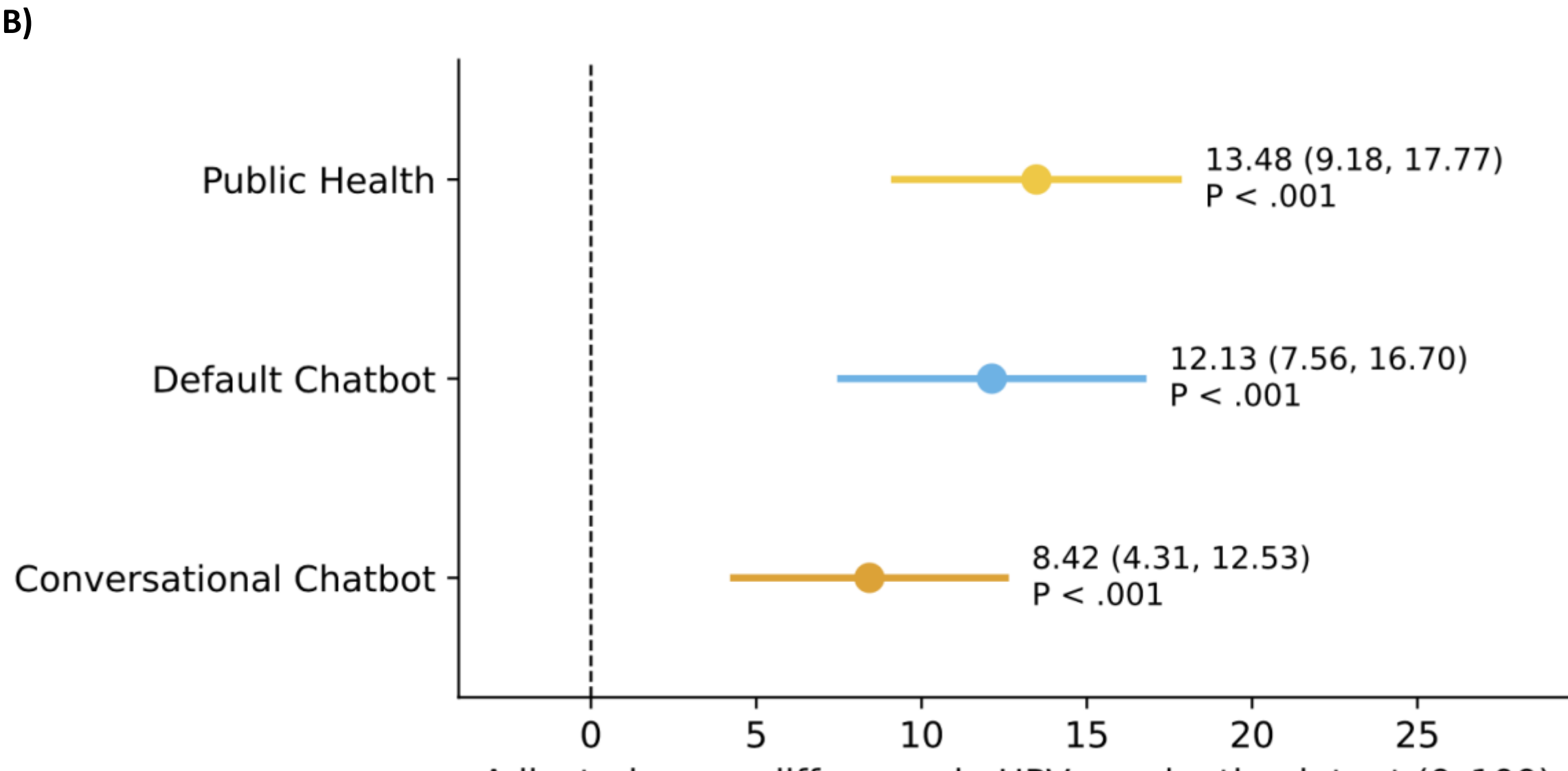

C)

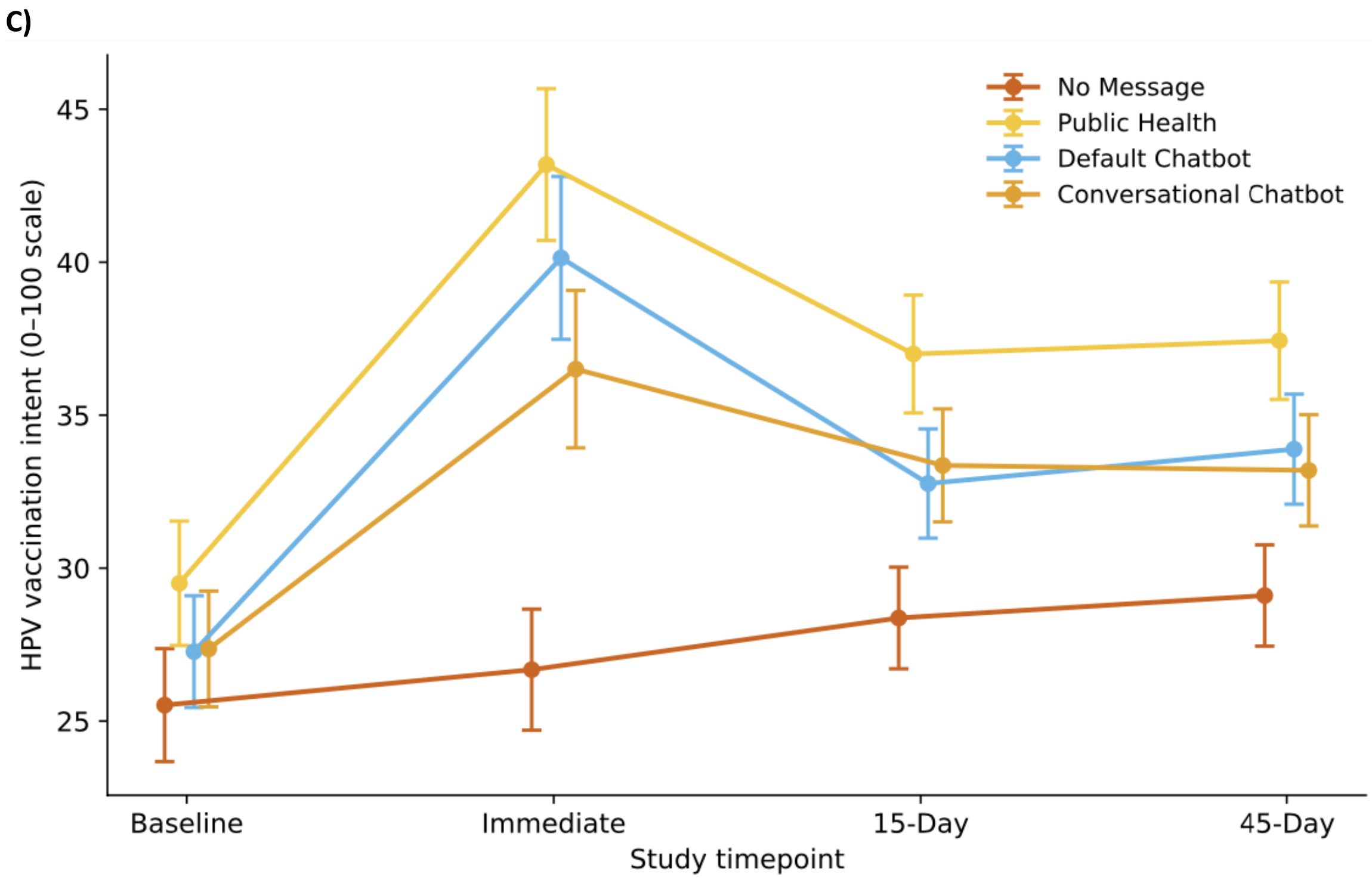

**Figure 1. Immediate change in parents' 12-month HPV-vaccination intention after each intervention (N = 1297).**

Panel A shows raw mean change scores (post- minus pre-intervention; 0–100 scale) for the four arms. Vertical lines display mean and standard errors. Vertical jitter is added for visibility and does not reflect a meaningful variable. Panel B displays HC2-robust 95 % confidence intervals for increase in vaccine intent relative to the no message control, derived from the OLS model in Table 2 (column 1). All three interventions significantly increased intention relative to the No-Message control. Panel C plots trajectory for each arm at baseline, immediately post-treatment, 15-day follow-up, and 45-day follow-up to illustrate trajectory over time, with SE.

**Table 1. Participant Characteristics by Arm**

| | Overall (n=1297) | Public Health Reading Arm (n=325) | Default Chatbot (n=324) | Conversational Chatbot (n=324) | No Message (n=324) |
|---|---|---|---|---|---|
| **Age, mean (SD)** | 42.84 (6.93) | 43.29 (6.88) | 43.27 (6.91) | 42.25 (6.73) | 42.57 (7.17) |
| **Race (%)[1]** | | | | | |
| White | 1043 (80.4) | 255 (78.5) | 248 (76.5) | 269 (83.0) | 271 (83.6) |
| Black | 129 (9.9) | 31 (9.5) | 41 (12.7) | 27 (8.3) | 30 (9.3) |
| Mixed | 48 (3.7) | 15 (4.6) | 18 (5.6) | 10 (3.1) | 5 (1.5) |
| Other | 38 (2.9) | 12 (3.7) | 9 (2.8) | 10 (3.1) | 7 (2.2) |
| Asian | 39 (3.0) | 12 (3.7) | 8 (2.5) | 8 (2.5) | 11 (3.4) |
| **Country (%)** | | | | | |
| United States | 825 (63.6) | 202 (62.2) | 211 (65.1) | 202 (62.3) | 210 (64.8) |
| United Kingdom | 431 (33.2) | 113 (34.8) | 97 (29.9) | 113 (34.9) | 108 (33.3) |
| Canada | 41 (3.2) | 10 (3.1) | 16 (4.9) | 9 (2.8) | 6 (1.9) |
| **Child HPV Vaccine Status** | | | | | |
| Unvaccinated | 1138 (87.7) | 288 (88.6) | 277 (85.5) | 284 (87.7) | 289 (89.2) |
| Unsure | 159 (12.3) | 37 (11.4) | 47 (14.5) | 40 (12.3) | 35 (10.8) |
| **Child Covid Vaccine Status** | | | | | |
| Unvaccinated | 967 (74.6) | 230 (70.8) | 235 (72.5) | 250 (77.2) | 252 (77.8) |
| At least 1 dose | 317 (24.4) | 92 (28.3) | 87 (26.9) | 71 (21.9) | 67 (20.7) |
| Unsure | 13 (1.0) | 3 (0.9) | 2 (0.6) | 3 (0.9) | 5 (1.5) |
| **Parent Vaccinated for COVID-19** | 673 (51.9) | 172 (52.9) | 171 (52.8) | 164 (50.6) | 166 (51.2) |
| **Political Ideology** | | | | | |
| Very Conservative | 106 (8.2) | 24 (7.4) | 26 (8.0) | 29 (9.0) | 27 (8.3) |
| Conservative | 227 (17.5) | 54 (16.6) | 49 (15.1) | 59 (18.2) | 65 (20.1) |
| Somewhat Conservative | 224 (17.3) | 53 (16.3) | 58 (17.9) | 54 (16.7) | 59 (18.2) |
| Moderate | 452 (34.8) | 118 (36.3) | 111 (34.3) | 113 (34.9) | 110 (34.0) |
| Somewhat Liberal | 143 (11.0) | 37 (11.4) | 43 (13.3) | 36 (11.1) | 27 (8.3) |
| Liberal | 111 (8.6) | 31 (9.5) | 30 (9.3) | 20 (6.2) | 30 (9.3) |
| Very Liberal | 34 (2.6) | 8 (2.5) | 7 (2.2) | 13 (4.0) | 6 (1.9) |
| **Religious** | | | | | |
| Not at all religious | 459 (35.4) | 114 (35.1) | 126 (38.9) | 110 (34.0) | 109 (33.6) |
| Slightly religious | 272 (21.0) | 70 (21.5) | 63 (19.4) | 72 (22.2) | 67 (20.7) |
| Moderately religious | 262 (20.2) | 64 (19.7) | 66 (20.4) | 73 (22.5) | 59 (18.2) |
| Very religious | 210 (16.2) | 50 (15.4) | 45 (13.9) | 47 (14.5) | 68 (21.0) |
| Extremely religious | 94 (7.2) | 27 (8.3) | 24 (7.4) | 22 (6.8) | 21 (6.5) |
| **Gender** | | | | | |
| Woman or female | 935 (72.1) | 230 (70.8) | 244 (75.3) | 232 (71.6) | 229 (70.7) |
| Man or male | 357 (27.5) | 94 (28.9) | 79 (24.4) | 89 (27.5) | 95 (29.3) |
| Other | 5 (0.4) | 1 (0.3) | 1 (0.3) | 3 (0.9) | 0 (0.0) |
| **Education** | | | | | |
| Less than high school | 10 (0.8) | 3 (0.9) | 1 (0.3) | 3 (0.9) | 3 (0.9) |
| High school diploma or GED | 202 (15.6) | 47 (14.5) | 56 (17.3) | 53 (16.4) | 46 (14.2) |
| Associate degree | 132 (10.2) | 30 (9.2) | 34 (10.5) | 35 (10.8) | 33 (10.2) |
| Some college, no degree | 340 (26.2) | 89 (27.4) | 81 (25.0) | 79 (24.4) | 91 (28.1) |
| Bachelor's degree | 426 (32.8) | 112 (34.5) | 108 (33.3) | 108 (33.3) | 98 (30.2) |
| Doctorate or professional degree | 32 (2.5) | 5 (1.5) | 8 (2.5) | 9 (2.8) | 10 (3.1) |
| Master's degree | 155 (12.0) | 39 (12.0) | 36 (11.1) | 37 (11.4) | 43 (13.3) |
| **Rural/Urban** | | | | | |
| Rural | 373 (28.8) | 88 (27.1) | 96 (29.6) | 93 (28.7) | 96 (29.6) |
| Suburb or Exurb | 607 (46.8) | 154 (47.4) | 141 (43.5) | 161 (49.7) | 151 (46.6) |
| Urban | 317 (24.4) | 83 (25.5) | 87 (26.9) | 70 (21.6) | 77 (23.8) |
| **Self-Reported Health** | | | | | |
| Poor | 32 (2.5) | 12 (3.7) | 10 (3.1) | 6 (1.9) | 4 (1.2) |
| Fair | 273 (21.0) | 62 (19.1) | 75 (23.1) | 73 (22.5) | 63 (19.4) |
| Good | 810 (62.5) | 210 (64.6) | 198 (61.1) | 188 (58.0) | 214 (66.0) |
| Excellent | 182 (14.0) | 41 (12.6) | 41 (12.7) | 57 (17.6) | 43 (13.3) |
| **Trust in Public Health Agency (1–5 scale; 5 = highest), mean (SD)** | 3.10 (1.11) | 3.13 (1.09) | 3.15 (1.08) | 3.08 (1.12) | 3.06 (1.16) |
| **Familiarity with AI** | | | | | |
| Not familiar at all | 38 (2.9) | 8 (2.5) | 8 (2.5) | 9 (2.8) | 13 (4.0) |
| Slightly familiar | 407 (31.4) | 109 (33.5) | 90 (27.8) | 108 (33.3) | 100 (30.9) |
| Moderately familiar | 428 (33.0) | 100 (30.8) | 124 (38.3) | 101 (31.2) | 103 (31.8) |
| Very familiar | 313 (24.1) | 79 (24.3) | 77 (23.8) | 77 (23.8) | 80 (24.7) |
| Extremely familiar | 111 (8.6) | 29 (8.9) | 25 (7.7) | 29 (9.0) | 28 (8.6) |
| **AI Usage** | | | | | |
| Never | 169 (13.0) | 51 (15.7) | 39 (12.0) | 45 (13.9) | 34 (10.5) |
| Rarely, once a month or less | 381 (29.4) | 91 (28.0) | 104 (32.1) | 89 (27.5) | 97 (29.9) |
| Occasionally, a few times a month | 364 (28.1) | 90 (27.7) | 88 (27.2) | 96 (29.6) | 90 (27.8) |
| Frequently, a few times a week | 267 (20.6) | 64 (19.7) | 67 (20.7) | 70 (21.6) | 66 (20.4) |
| Daily | 116 (8.9) | 29 (8.9) | 26 (8.0) | 24 (7.4) | 37 (11.4) |
| **AI Trust** | | | | | |
| Not at all | 89 (6.9) | 21 (6.5) | 24 (7.4) | 26 (8.0) | 18 (5.6) |
| Slightly | 317 (24.4) | 71 (21.8) | 88 (27.2) | 79 (24.4) | 79 (24.4) |
| Somewhat | 612 (47.2) | 175 (53.8) | 147 (45.4) | 142 (43.8) | 148 (45.7) |
| Very much | 269 (20.7) | 53 (16.3) | 64 (19.8) | 74 (22.8) | 78 (24.1) |
| Completely | 10 (0.8) | 5 (1.5) | 1 (0.3) | 3 (0.9) | 1 (0.3) |
| **Likelihood to vaccinate child for COVID-19 in the next 12 months, mean (SD)** | 9.92 (21.88) | 12.05 (25.32) | 10.82 (22.24) | 8.49 (20.60) | 8.30 (18.67) |
| **Likelihood to vaccinate child for flu in the** | 27.30 (36.40) | 27.54 (36.57) | 26.88 (35.41) | 27.34 (37.14) | 27.46 (36.63) |
| **next 12 months, mean (SD)** | | | | | |
| **Likelihood to vaccinate child for HPV in the next 12 months, mean (SD)** | 27.42 (34.23) | 29.50 (36.65) | 27.27 (32.88) | 27.36 (34.06) | 25.52 (33.23) |
| **PACV-5 (General Parental Vaccine Hesitancy Scale)** | 3.43 (1.03) | 3.43 (1.00) | 3.50 (1.08) | 3.38 (1.01) | 3.42 (1.03) |
| **Participants Recruited via Prolific** | 1185 (91.4) | 297 (91.4) | 296 (91.4) | 296 (91.4) | 296 (91.4) |
| **Reasons for HPV Non-Vaccination** | | | | | |
| the vaccine not being recommended | 41.55 (32.95) | 40.91 (32.83) | 43.42 (33.10) | 40.32 (33.89) | 41.51 (32.10) |
| the vaccine not being necessary | 57.81 (34.28) | 56.78 (35.21) | 58.37 (33.78) | 58.97 (34.23) | 57.14 (34.00) |
| their child being fearful | 27.92 (31.35) | 25.39 (30.10) | 30.51 (32.60) | 29.77 (31.47) | 26.23 (31.13) |
| their child not being sexually active | 65.22 (38.02) | 63.76 (38.89) | 66.20 (37.39) | 64.42 (37.19) | 66.52 (38.67) |
| their child not being old enough | 49.55 (36.79) | 46.07 (36.29) | 51.22 (35.82) | 50.29 (36.89) | 50.64 (38.04) |
| side effects | 70.37 (32.74) | 69.98 (33.72) | 70.74 (33.41) | 71.44 (32.05) | 69.34 (31.91) |
| it being a family decision | 68.47 (31.71) | 67.50 (32.47) | 66.67 (33.65) | 69.65 (30.61) | 69.94 (30.14) |
| it not being a school requirement | 46.97 (36.49) | 51.69 (37.17) | 44.97 (35.57) | 44.76 (36.53) | 46.49 (36.45) |
| needing more information | 58.88 (36.24) | 59.03 (36.92) | 61.34 (34.79) | 56.19 (36.72) | 58.97 (36.48) |
| not believing in vaccinations | 31.83 (33.49) | 34.32 (34.85) | 31.69 (32.79) | 29.60 (32.67) | 31.80 (33.67) |
| the vaccine being too new | 48.82 (34.98) | 49.14 (35.03) | 52.87 (34.91) | 46.22 (35.17) | 47.18 (34.63) |
| letting their child decide when they're older | 56.41 (33.71) | 55.31 (34.82) | 59.62 (33.04) | 53.70 (33.91) | 57.01 (32.90) |
| other | 65.56 (34.25) | 70.51 (30.62) | 65.68 (34.53) | 66.40 (35.23) | 58.38 (36.62) |

[1]Racial categories follow the five-category classification recommended by the UK Office for National Statistics, as used by Prolific; Connect panel participants were mapped to the closest equivalent category. "Other" race/ethnicity includes participants who selected "Other" on the Prolific recruitment platform (n=36) or "Other" or "Prefer not to say" in the Connect panel (n=2).

**Table 2. Results from Ordinary Least Squares Regression**

| Outcome | Contrast | Estimate | 95% CI | p-value | Cohen's d | d 95% CI |
|---|---|---|---|---|---|---|
| Immediate HPV vaccine likelihood (0-100) | | | | | | |
| | Public Health vs No Message | 13.48 | 9.18 to 17.77 | <0.001 | 0.53 | 0.36 to 0.70 |
| | Default Chatbot vs No Message | 12.13 | 7.56 to 16.70 | <0.001 | 0.48 | 0.30 to 0.65 |
| | Conversational Chatbot vs No Message | 8.42 | 4.31 to 12.53 | <0.001 | 0.33 | 0.17 to 0.49 |
| | Default Chatbot vs Public Health | -1.35 | -5.94 to 3.24 | 0.564 | -0.05 | -0.23 to 0.13 |
| | Conversational Chatbot vs Public Health | -5.05 | -9.41 to -0.69 | 0.023 | -0.2 | -0.37 to -0.03 |
| Immediate COVID vaccine likelihood (0-100) | | | | | | |
| | Public Health vs No Message | -0.3 | -2.73 to 2.12 | 0.806 | -0.01 | -0.12 to 0.10 |
| | Default Chatbot vs No Message | -1.51 | -4.04 to 1.02 | 0.241 | -0.07 | -0.18 to 0.05 |
| | Conversational Chatbot vs No Message | -1.34 | -3.65 to 0.97 | 0.256 | -0.06 | -0.17 to 0.04 |
| | Default Chatbot vs Public Health | -1.21 | -3.82 to 1.40 | 0.363 | -0.06 | -0.17 to 0.06 |
| | Conversational Chatbot vs Public Health | -1.04 | -3.50 to 1.43 | 0.41 | -0.05 | -0.16 to 0.07 |
| Immediate flu vaccine likelihood (0-100) | | | | | | |
| | Public Health vs No Message | -0.92 | -4.26 to 2.42 | 0.59 | -0.03 | -0.12 to 0.07 |
| | Default Chatbot vs No Message | 2.09 | -1.72 to 5.90 | 0.282 | 0.06 | -0.05 to 0.16 |
| | Conversational Chatbot vs No Message | -0.4 | -3.85 to 3.04 | 0.818 | -0.01 | -0.11 to 0.08 |
| | Default Chatbot vs Public Health | 3.01 | -0.57 to 6.59 | 0.1 | 0.08 | -0.02 to 0.18 |
| | Conversational Chatbot vs Public Health | 0.51 | -2.81 to 3.84 | 0.762 | 0.01 | -0.08 to 0.11 |

| Immediate PACV-5 (0-6) | | | | | | |
|---|---|---|---|---|---|---|
| | Public Health vs No Message | 0.01 | -0.15 to 0.17 | 0.891 | 0.01 | -0.14 to 0.16 |
| | Default Chatbot vs No Message | 0.1 | -0.08 to 0.29 | 0.261 | 0.1 | -0.08 to 0.28 |
| | Conversational Chatbot vs No Message | -0.03 | -0.20 to 0.15 | 0.772 | -0.02 | -0.19 to 0.14 |
| | Default Chatbot vs Public Health | 0.09 | -0.07 to 0.26 | 0.269 | 0.09 | -0.07 to 0.25 |
| | Conversational Chatbot vs Public Health | -0.04 | -0.20 to 0.12 | 0.653 | -0.04 | -0.19 to 0.12 |
| 15-day HPV vaccine likelihood (0-100) | | | | | | |
| | Public Health vs No Message | 6.1 | 2.35 to 9.84 | 0.001 | 0.18 | 0.07 to 0.29 |
| | Default Chatbot vs No Message | 3.28 | -0.35 to 6.91 | 0.076 | 0.1 | -0.01 to 0.20 |
| | Conversational Chatbot vs No Message | 3.82 | 0.31 to 7.33 | 0.033 | 0.11 | 0.01 to 0.21 |
| | Default Chatbot vs Public Health | -2.81 | -6.74 to 1.11 | 0.16 | -0.08 | -0.20 to 0.03 |
| | Conversational Chatbot vs Public Health | -2.28 | -6.09 to 1.54 | 0.243 | -0.07 | -0.18 to 0.05 |
| 45-day HPV vaccine likelihood (0-100) | | | | | | |
| | Public Health vs No Message | 5.88 | 2.09 to 9.67 | 0.002 | 0.17 | 0.06 to 0.28 |
| | Default Chatbot vs No Message | 3.71 | -0.05 to 7.46 | 0.053 | 0.11 | -0.00 to 0.22 |
| | Conversational Chatbot vs No Message | 2.96 | -0.66 to 6.58 | 0.109 | 0.09 | -0.02 to 0.19 |
| | Default Chatbot vs Public Health | -2.17 | -6.14 to 1.80 | 0.283 | -0.06 | -0.18 to 0.05 |
| | Conversational Chatbot vs Public Health | -2.92 | -6.76 to 0.93 | 0.137 | -0.09 | -0.20 to 0.03 |

**Table 3. Results from Logistic Regression**

| Outcome | Contrast | OR | OR 95% CI | p-value |
|---|---|---|---|---|
| Interested in vaccine follow-up material | | | | |
| | Public Health vs No Message | 0.99 | 0.57 to 1.71 | 0.97 |
| | Default Chatbot vs No Message | 1.51 | 0.75 to 3.01 | 0.246 |
| | Conversational Chatbot vs No Message | 1.01 | 0.56 to 1.80 | 0.984 |
| | Default Chatbot vs Public Health | 1.52 | 0.79 to 2.93 | 0.21 |
| | Conversational Chatbot vs Public Health | 1.02 | 0.58 to 1.80 | 0.954 |
| Clicked vaccine appointment link (US only)[1] | | | | |
| | Public Health vs No Message | 0 | 0.00 to Inf | 0.999 |
| | Default Chatbot vs No Message | 2.69 | 0.28 to 26.29 | 0.395 |
| | Conversational Chatbot vs No Message | 3.06 | 0.31 to 29.91 | 0.337 |
| | Default Chatbot vs Public Health | Inf | 0.00 to Inf | 0.999 |
| | Conversational Chatbot vs Public Health | Inf | 0.00 to Inf | 0.999 |
| Donated to immunization charity | | | | |
| | Public Health vs No Message | 1.27 | 0.80 to 2.03 | 0.317 |
| | Default Chatbot vs No Message | 1.31 | 0.80 to 2.14 | 0.289 |
| | Conversational Chatbot vs No Message | 1.11 | 0.68 to 1.81 | 0.669 |
| | Default Chatbot vs Public Health | 1.03 | 0.65 to 1.63 | 0.903 |
| | Conversational Chatbot vs Public Health | 0.88 | 0.55 to 1.40 | 0.578 |
| Vaccinated by 15 days | | | | |
| | Public Health vs No Message | 1.48 | 0.70 to 3.14 | 0.305 |
| | Default Chatbot vs No Message | 1.88 | 0.90 to 3.96 | 0.095 |
| | Conversational Chatbot vs No Message | 1.17 | 0.52 to 2.64 | 0.699 |
| | Default Chatbot vs Public Health | 1.27 | 0.67 to 2.42 | 0.463 |
| | Conversational Chatbot vs Public Health | 0.79 | 0.39 to 1.61 | 0.518 |
| Vaccinated by 45 days | | | | |
| | Public Health vs No Message | 1.57 | 0.82 to 3.03 | 0.174 |
| | Default Chatbot vs No Message | 1.72 | 0.89 to 3.31 | 0.106 |
| | Conversational Chatbot vs No Message | 1.14 | 0.56 to 2.29 | 0.721 |
| | Default Chatbot vs Public Health | 1.09 | 0.62 to 1.92 | 0.764 |
| | Conversational Chatbot vs Public Health | 0.72 | 0.39 to 1.33 | 0.293 |

[1]The vaccine appointment link outcome was available to US participants only (n=825). Zero participants in the Public Health arm clicked the link, producing complete separation in logistic regression and unstable odds ratio estimates (0 or ∞) with uninformative confidence intervals.

# Supplementary Information

## Supplementary Methods Details

### Eligibility Screening and Exclusion Criteria

Eligibility was determined via a prescreening survey assessing parental status, child age, and HPV vaccination status. Participants were excluded from analysis if they failed attention checks, were flagged as likely bots based on an embedded invisible reCAPTCHA score below 0.5, or provided inconsistent initials identifying their youngest HPV vaccine–eligible child across surveys. Participants who no longer reported having an HPV-eligible child at Survey 2 were excluded prior to randomization.

### Power Calculation

The preregistered sample size was based on an a priori power calculation conducted in G*Power, assuming 95% power, a two-sided α of .05, and a small standardized effect size (d = 0.10), yielding a target of 431 participants per arm. Due to recruitment constraints, the final analytic sample included 1297 participants across four arms.

### Randomization

Randomization was conducted in Python using a reproducible pseudorandom shuffle of eligible participants from Survey 1 with a fixed random seed (random_state=42). After shuffling, participants were assigned in a 1:1:1:1 ratio to the 4 study arms by dividing the randomized list into 4 approximately equal groups. Randomization was not stratified or blocked. Allocation was prespecified by participant ID before Survey 2 and applied using the online study platform; study staff did not determine assignment at the point of intervention delivery.

### Conversational and Persuasion Strategy Analyses

To characterize chatbot conversations, we conducted exploratory content analyses among the complete-case sample (n=930) using AI-assisted annotation procedures described in prior work.[1,2] GPT-4o was prompted to generate candidate persuasion strategies, which were deduplicated into a final set of 23 strategies. GPT-4o-mini rated the prevalence of each strategy at the sentence level using an ordinal scale. Principal component analysis with varimax rotation was used to identify higher-order components, and LASSO regression was employed as a robustness check. Comparable strategy coding was applied to the public health materials.

Separately, GPT-4o-mini rated chatbot adherence to motivational interviewing techniques, including open-ended questions, affirmation, reflective listening, summarizing, and use of an elicit–share–elicit structure. Linear regressions examined associations between these techniques and post-intervention vaccination intent.

All prompts used by GPT to annotate conversations appear in Table S11.

### Clinical Review of Chatbot Responses

To assess factual accuracy and clinical appropriateness, a licensed clinician reviewed a random sample of 50 chatbot responses classified as high in evidence-based information. All reviewed responses were deemed factually accurate and clinically appropriate.

## Supplementary Results: Exploratory Analyses of Chatbot Content and Persuasion Strategies

### Identification and Prevalence of Persuasion Strategies

Using a multi-step content analysis procedure previously, we identified 23 candidate persuasion strategies (Table S12). "Evidence-Based Information" and "Listen Actively" had the highest average levels of usage across the chatbot conversations, while “Personal Stories” had the lowest (Figure S6, Table S12). The two chatbots differed significantly in their usage of several strategies. For instance, the default chatbot used “Discuss Disease Prevalence” more frequently ($p<.001$) while the conversational chatbot used “Long Term Health Investment” more often ($p<.05$) (Table S12). As expected, the static public health material had very high levels of "Evidence-Based Information" and no usage of conversational strategies such as "Listen Actively" or "Opportunity for questions" (Table S12).

**Principal Component Analysis**

Six PCA components explained 62.8% of the conversation-level variance in the 23 persuasion strategy prevalence scores across the 449 chatbot dialogs: Open Dialogue and Support; Health Benefit Framing vs. Resources and Trusted Experts; Age and Timing Emphasis; Rational Planning vs. Empathetic Personalization; Evidence-Based Information vs. Personalized Empathy; and Herd Immunity vs. Safety Concerns (Figure S7). Components 2, 4, 5, and 6 are bipolar, reflecting contrasting messaging strategies along a shared thematic axis (e.g. benefit framing vs. authority/resources; collective benefits vs. individual-level risk). Labels reflect manual interpretation of high-loading strategies. In a linear regression predicting post-intervention vaccination intent, higher scores on the Resources and Trusted Experts pole of Component 2 (i.e. a negative coefficient on the bipolar scale), the Evidence-Based Information pole of Component 5, and the Herd Immunity pole of Component 6 were associated with higher intent, whereas the opposing poles (Health Benefit Framing, Personalized Empathy, and Safety Concerns) were associated with lower intent (Table S13).

**LASSO Regression**

LASSO regression ($\lambda = 0.18$) retained 22 of the 23 individual persuasion strategies and broadly aligned with the PCA results, though some differences emerged at the level of individual features (Table S14). Consistent with the component-based analyses, strategies emphasizing external validation and collective benefit (e.g. Involve Trusted Experts and Discuss Herd Immunity) were positively associated with higher post-intervention vaccination intent, whereas strategies emphasizing personalized empathy (e.g. Listen Actively and Personalized Approach) were negatively associated.

**Motivational Interviewing Adherence**

Across both chatbots, adherence to motivational interviewing, a validated approach to addressing vaccine hesitancy, was high, as assessed by GPT-4o-mini.[3-7] Among the five individual techniques of motivational adherence, there was high adherence to sharing knowledge, moderate adherence to affirmation, open-ended questions, and reflective listening, but low adherence to summarizing (Table S15). The default chatbot demonstrated higher adherence to motivational interviewing techniques relative to the conversational chatbot (Table S15). Affirmation and Sharing Knowledge techniques were both positively associated with post-intervention vaccination intent in linear regression analyses (Table S16).

**Supplement References**

1. T. H. Costello, G. Pennycook, D. G. Rand, Durably reducing conspiracy beliefs through dialogues with AI. Science 385, eadq1814 (2024).

2. T. H. Costello, G. Pennycook, D. Rand, Just the facts: How dialogues with AI reduce conspiracy beliefs. [Preprint] (2025). Available at: https://osf.io/h7n8u_v1 [Accessed 14 April 2025].

3. Communicating with caregivers about the Human Papillomavirus vaccination: facilitator’s guide. Available at: https://www.who.int/publications/i/item/WHO-EURO-2023-6839-46605-67665 [Accessed 14 April 2025].

4. P. H. A. of Canada, Motivational interviewing: A powerful tool to address vaccine hesitancy, CCDR 46(4). (2020). Available at: https://www.canada.ca/en/public-health/services/reports-publications/canada-

communicable-disease-report-ccdr/monthly-issue/2020-46/issue-4-april-2-2020/article-6-canvax-addressing-vaccine-hesitancy.html [Accessed 14 April 2025].

5. M. Galeshi, H. Shirafkan, S. Yazdani, Z. Motaghi, Effectiveness of motivational interview on psychological distress of women with human papilloma virus: a randomized clinical trial. BMC Women's Health 25, 73 (2025).

6. D. N. S. Chan, K. C. Choi, P. P. K. Lee, W. K. W. So, The Effects of Motivational Interviewing on Promoting Human Papillomavirus Vaccination Initiation and Completion Among South Asian Mother/Daughter Dyads: A Pilot Randomised Controlled Trial. Int.J. Behav. Med. (2025). https://doi.org/10.1007/s12529-025-10349-y.

7. J. E. Reno, et al., Improving Provider Communication about HPV Vaccines for Vaccine-Hesitant Parents Through the Use of Motivational Interviewing. Journal of Health Communication 23, 313–320 (2018).

**Public Health Reading Condition Stimuli**

Canadian Material

# Human papillomavirus / HPV:

# Spread, prevention and risks

## How HPV spreads

HPV is the most common sexually transmitted infection in Canada and around the world. It spreads by skin-to-skin contact and is easily spread through sexual touching and during vaginal, anal or oral sex.

HPV infects the skin and mucous membrane, which is the moist, inner lining of body parts like the:

- anus
- mouth
- vagina
- rectum

It's possible to be infected by more than 1 type of HPV at a time. It's also possible to be infected again with the same HPV type you previously had.

In rare cases, some types may cause a condition where warts develop on the respiratory tract.

If you're pregnant, in very rare cases, HPV can spread from you to your baby during pregnancy or delivery.

You can spread HPV to a sexual partner even if you don't have symptoms.

## Who is at risk

Anyone who is sexually active is at risk of HPV infections, regardless of their sex, gender identity or sexual
orientation. People who have received an HPV vaccine are at much lower risk of HPV

infections.

## Preventing HPV

### HPV vaccination

HPV vaccination is the most effective way to protect yourself against HPV infections and the cancers that this virus can cause.

The 9vHPV vaccine is recommended as it provides protection against the greatest number of HPV types and associated diseases.

The 9vHPV vaccine provides protection against 7 high-risk HPV types, accounting for 90% of all cervical cancers. This vaccine also protects against 2 low-risk HPV types, which:

- cause about 90% of anogenital warts (relating to the anus and genitals)
- can cause a rare but serious condition where warts develop on the respiratory tract

### Recommendations for vaccination

The 9vHPV vaccine is approved in Canada for individuals aged 9 to 45 years. It's best to be vaccinated at a younger age before exposure to HPV. Talk to your health care provider to find out how many doses are recommended for you.

The HPV vaccine is offered as part of routine vaccinations through school-based programs, public health clinics or catch-up programs. Provinces and territories may extend the eligibility period for publicly funded vaccines:

- until the completion of high school
- beyond high school for some adults

Every province and territory in Canada has a publicly funded HPV vaccine program. HPV vaccine is also available for private purchase for individuals who aren't included in a publicly funded HPV vaccination program. Talk to a health care provider or your local public health unit for more information.

### Vaccine safety and effectiveness

HPV vaccines work very well. Studies have shown they're highly effective at

preventing:

- HPV infections
- HPV-related cancers
- anogenital warts

HPV vaccines are safe. Experts from around the world have:

- been monitoring their safety since before they were first made available
- not identified any safety concerns

### Importance of vaccination

Canada aims to eliminate cervical cancer by 2040 through improved HPV vaccination, screening and follow-up. In 2021, 81% of 14-year-old males and 86% of 14-year-old females had received at least 1 dose of HPV vaccine.

### Cervical cancer screening

Cervical cancer screening can detect pre-cancerous lesions. Getting regularly tested for cervical abnormalities can help protect against cervical cancer.

### Condom use

Condoms do not eliminate the risk of HPV infections. This is because there can still be skin-to-skin contact in areas the condom doesn't cover. However, proper use of condoms and dental dams can reduce the chances of:

- getting HPV
- passing HPV to a sexual partner

Using a condom is very effective at helping to protect against many other sexually transmitted infections.

# Reasons to Get Vaccinated

## Key points

- All children ages 11–12-years should get HPV vaccine to protect against cancers caused by HPV infections.

## Why it's important

Almost every unvaccinated person who is sexually active will get HPV at some time in their life. About 13 million Americans, including teens, become infected with HPV each year. Most HPV infections will go away on their own. But infections that don't go away can cause certain types of cancer.

HPV can cause cancers of the:

- Cervix, vagina, and vulva in women
- Penis in men
- Anus in both women and men
- Back of the throat (called oropharyngeal cancer), including the base of the tongue and tonsils, in both men and women

## What the data show

HPV vaccination works.

HPV infections, genital warts, and cervical pre-cancers (abnormal cells on the cervix that can lead to cancer) have dropped since the vaccine has been in use in the United States.

- Infections with HPV types that cause most HPV cancers and genital warts have dropped 88% among teen girls and 81% among young adult women.
- Among vaccinated women, the percentage of cervical pre-cancers caused by the HPV types most often linked to cervical cancer have dropped by 40 percent.

HPV vaccination is cancer prevention.

HPV is estimated to cause nearly 36,500 cases of cancer in men and women every year in the United States. HPV vaccination can prevent 90% of these cancers by preventing the infections that cause them.

Preventing cancer is better than treating it.

HPV can cause several kinds of cancer. Cervical cancer can be detected early with a screening test. Other types of cancer caused by HPV may not be detected until they are more serious. HPV vaccination prevents infections that cause these cancers.

Early protection works best.

Most children only need two doses of HPV vaccine when vaccinated before age 15 years. You can take advantage of any visit to your child's doctor to get recommended vaccines for your child:

- Annual checkups
- Sports physicals
- Annual flu shot or other vaccine visits

HPV vaccination provides safe, effective, and long-lasting protection.

With more than 135 million doses distributed in the United States, HPV vaccine has a reassuring safety record that's backed by over 15 years of monitoring and research.

# Impact of the HPV Vaccine

HPV vaccination provides safe, effective, and long-lasting protection against cancers caused by HPV.

## HPV vaccination is very safe

Over 15 years of monitoring and research during the vaccination program have continued to show that HPV vaccination is safe.

More than 135 million doses of HPV vaccines have been distributed since they were licensed. Data continue to show the vaccines are safe and effective.

As with all approved vaccines, CDC and FDA closely monitor the safety of HPV vaccines. Any detected safety concerns are reported to health officials, healthcare professionals, and the public.

### HPV vaccine doesn't cause fertility problems

Not getting HPV vaccine leaves people at risk for HPV cancers and precancers (abnormal cells that can lead to cancer).

- People who develop a cancer caused by HPV will need treatment that can sometimes limit their ability to have children, such as a hysterectomy (for women), chemotherapy, or radiation.
- Treatment for cervical precancer could also put women at risk for problems with their cervix, which can sometimes cause preterm delivery.

## HPV vaccination prevents cancer-causing infections & precancers

HPV infections and cervical precancers have dropped since 2006, when HPV vaccines were first used in the United States.

- Among teen girls, infections with HPV types that cause most HPV cancers and genital warts have dropped 88 percent.
- Among young adult women, infections with HPV types that cause most HPV cancers and genital warts have dropped 81 percent.
- Among vaccinated women, the percentage of cervical precancers caused by the HPV types most often linked to cervical cancer has dropped by 40 percent.

## Long-lasting protection

The protection provided by HPV vaccines lasts a long time. People who received HPV vaccines were followed for at least about 12 years, and their protection against HPV has remained high with no evidence of decreasing over time.

# HPV vaccine

The HPV vaccine helps protect against human papillomavirus (HPV). It's recommended for children aged 12 to 13 years old and people at higher risk from HPV.

## What the HPV vaccine is for

The HPV vaccine reduces your chances of getting human papillomavirus (HPV), a common virus that's spread through skin contact (usually when having sex).

Most types of HPV are harmless. But some types are linked to an increased risk of certain types of cancer, including:

- cervical cancer
- mouth cancer
- anal cancer
- penile cancer
- vulval cancer
- vaginal cancer

HPV can also cause genital warts.

## Who should have the HPV vaccine

The HPV vaccine is recommended for children aged 12 to 13 years old and people at higher risk from HPV.

### Children aged 12 to 13

All children aged 12 to 13 (school year 8) are offered the HPV vaccine.

If you missed getting vaccinated when you were 12 or 13, the HPV vaccine is available for free on the NHS for:

- all girls under 25
- boys born after 1 September 2006

Contact your school nurse, school vaccination team or GP surgery if you or your child were eligible for the HPV vaccine but did not get vaccinated.

## Who cannot have the HPV vaccine

Most people who are eligible for the HPV vaccine can have it.

You only cannot have the vaccine if you've had a serious allergic reaction (anaphylaxis) to a previous dose of the vaccine, or an ingredient in the vaccine.

There's no evidence the vaccine is harmful if you're pregnant, but sometimes you may be advised to wait until you're no longer pregnant before having the vaccine.

You can have the HPV vaccine while breastfeeding.

## Getting vaccinated if you're unwell

If you have a high temperature or feel too unwell to do your normal activities, wait until you're feeling better before having the vaccine.

## How to get the HPV vaccine

There are different ways to get the HPV vaccine.

| Who should have it | Where to get it |
|---|---|
| Children aged 12 to 13 (school year 8) | Secondary school (or community clinics for those not in school) |
| Girls under 25 and boys born after 1 September 2006 who missed having the vaccine at school | Check with your school nurse, school vaccination team or GP surgery |
| Men aged 45 and under who have sex with men, and other people at higher risk of HPV | Sexual health clinics or HIV clinics |

## How the HPV vaccine is given

The HPV vaccine is given as an injection into your arm.

The number of doses you need depends on your age and how well your immune system works:

- people under 25 usually have 1 dose
- people aged 25 to 45 usually have 2 doses (given between 6 months and 2 years apart)
- people with a weakened immune system have 3 doses (given over a 12-month period)

Side effects of the HPV vaccine

The most common side effects of the HPV vaccine are mild and do not last long.

They can include:

- swelling or pain where the injection was given
- a headache
- a high temperature
- dizziness
- feeling sick
- tiredness

More serious side effects such as a severe allergic reaction are very rare. The person who vaccinates you will be trained to deal with allergic reactions and treat them immediately.

How well the HPV vaccine works

The HPV vaccine works very well in reducing your risk of getting HPV.

Since the vaccine has been used, there has been a big drop in the number of young people getting conditions linked to HPV, such as cervical cancer and genital warts.

Research suggests that over time the HPV vaccine will help save thousands of lives in the UK.

Important

It's still important to attend cervical screening appointments if you've been vaccinated against HPV, as there's still a small chance you could get cervical cancer.

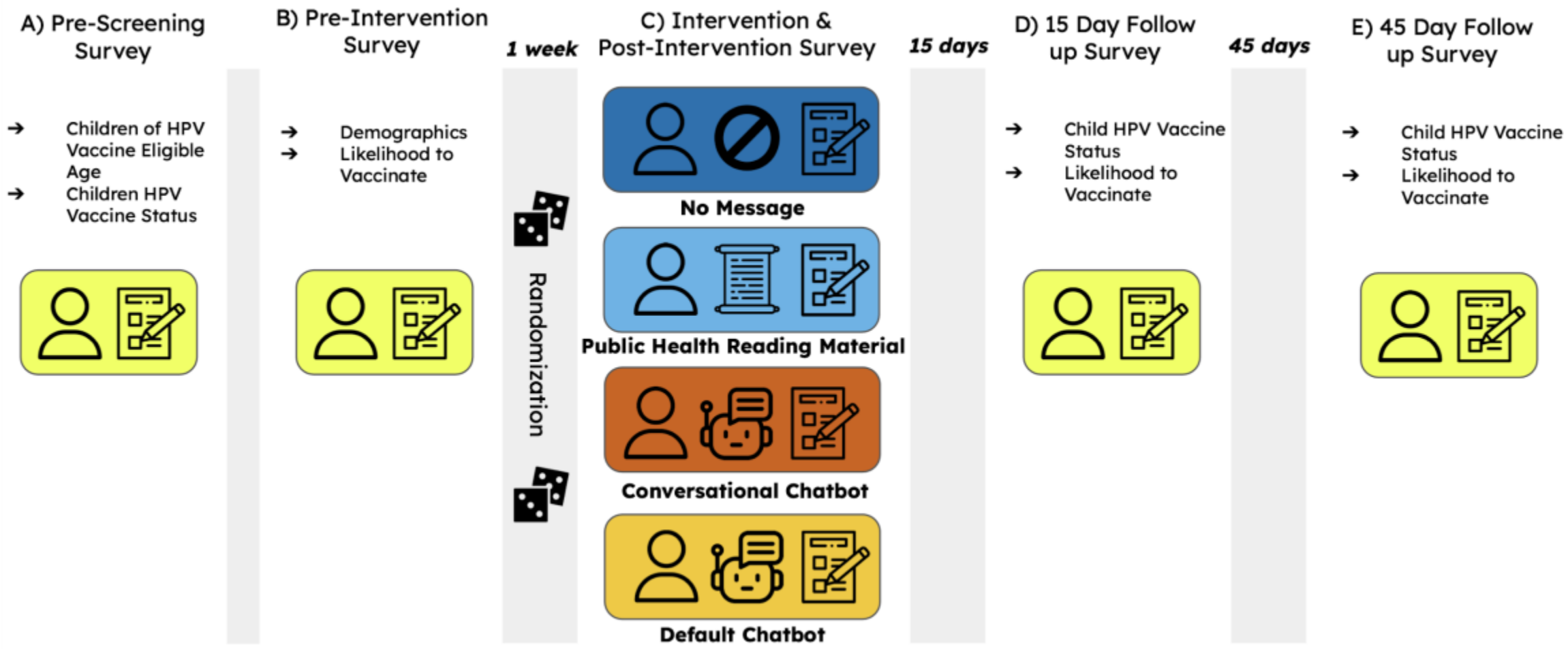

**Figure S1. Study Flow**

You will participate in a **conversation with an AI chatbot** about some of the **topics and ideas that you have already answered questions about** for a **3-minute** conversation.

The purpose of this dialogue is to see how humans and AI can engage around complicated topics. Please be open and honest in your responses. Remember that AI is neutral and nonjudgmental, and your participation is confidential.

The conversation will last **3 minutes** before you can proceed. Please take your time to engage thoughtfully.

When the conversation begins, you should see an AI avatar with chat bubbles "..." on the top left of your screen indicating it's generating responses. It can sometimes take up to 30s. If you don't see any icons or if it's taking too long to generate responses, try refreshing the page. If you run into further issues, please let us know.

When the conversation is over, you should see this message at the bottom: Scroll down and proceed to the next section.

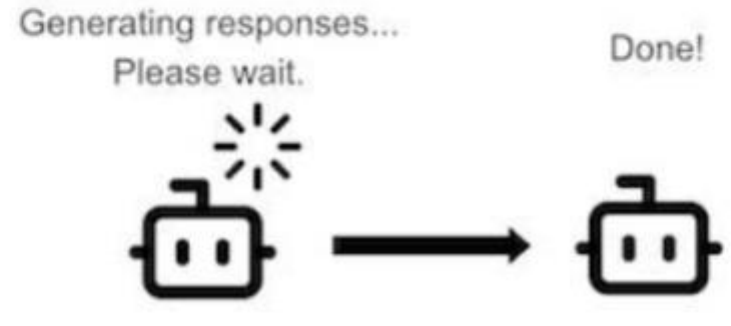

**Figure S2. Screen shown to participants before chatbot conversation.**

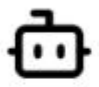

**Figure S3. Chatbot Interface.**
The agent itself was implemented via the OpenAI Chat-Completions API embedded in a Qualtrics Vegapunk widget: on every turn the interface sent the cumulative dialogue history plus the fixed system prompt to the API, and the assistant's text was rendered verbatim

**Figure S4. CONSORT Diagram of Participants.**

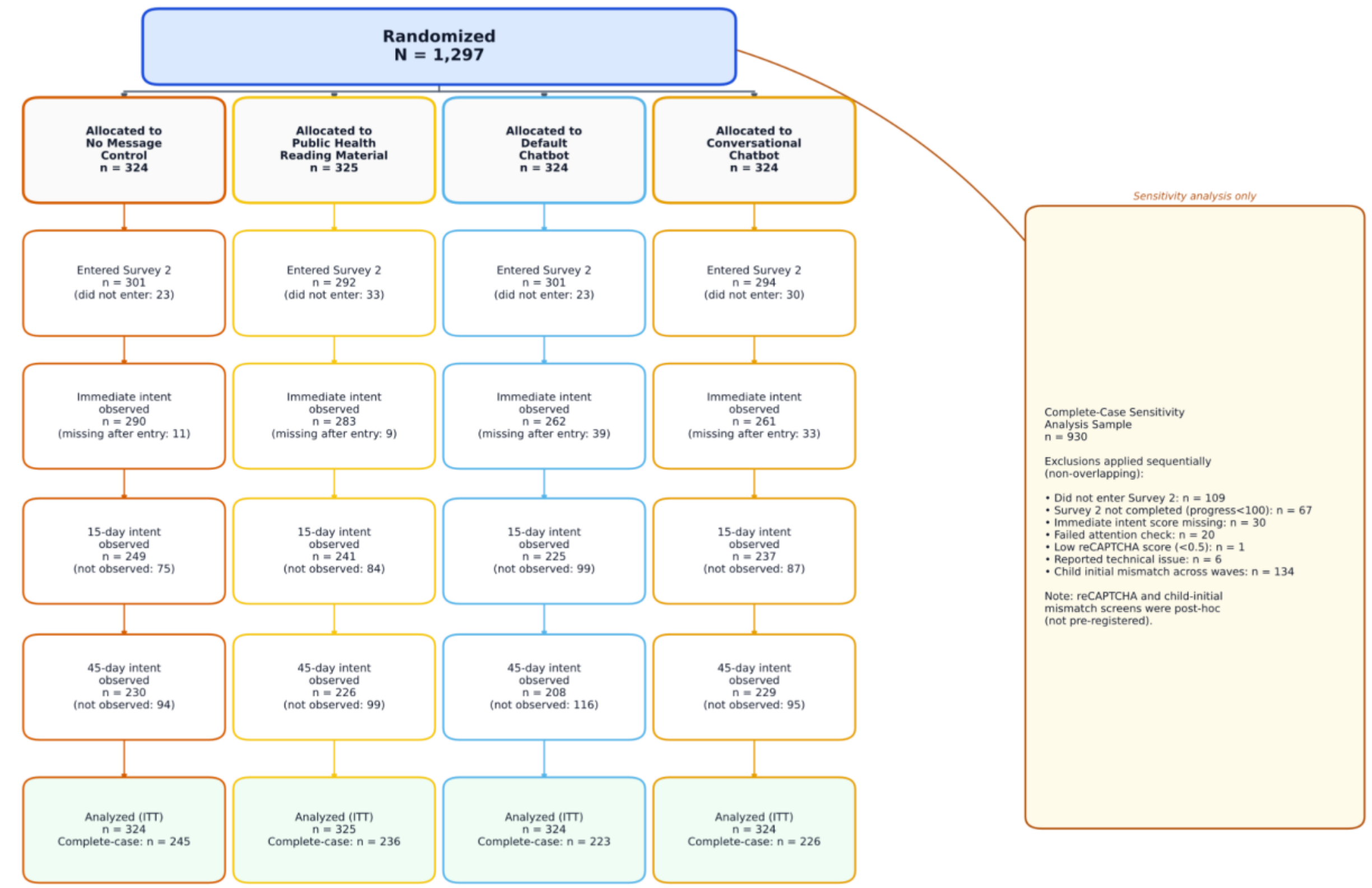

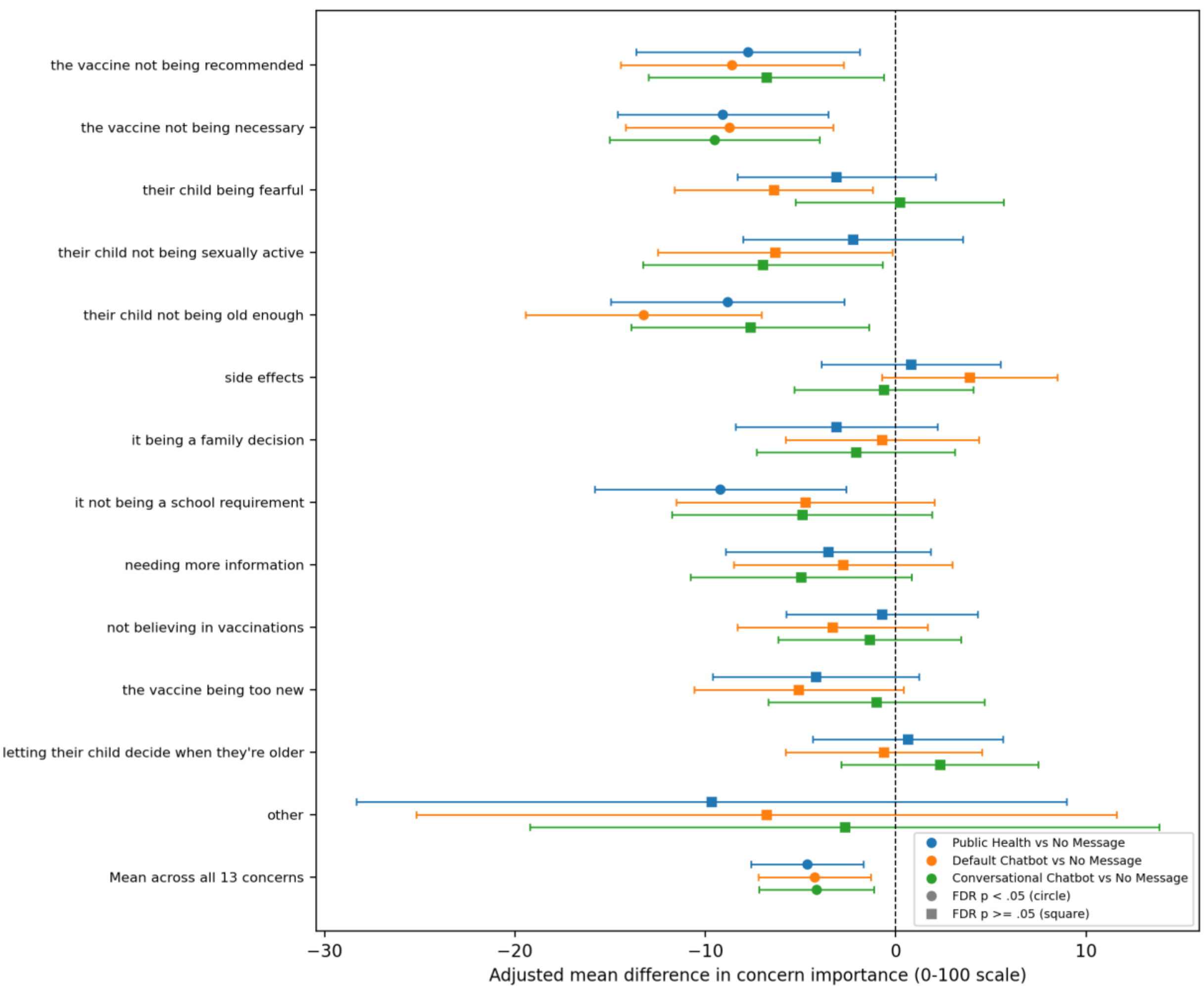

**Figure S5. Intervention effects on 13 concerns for HPV-vaccine hesitancy.** Coefficient plot (point estimates ± 95 % CI) from separate OLS models for importance of specific concerns (scale 0: Extremely Unimportant to 100: Extremely Important); estimates are relative to the No-Message control.

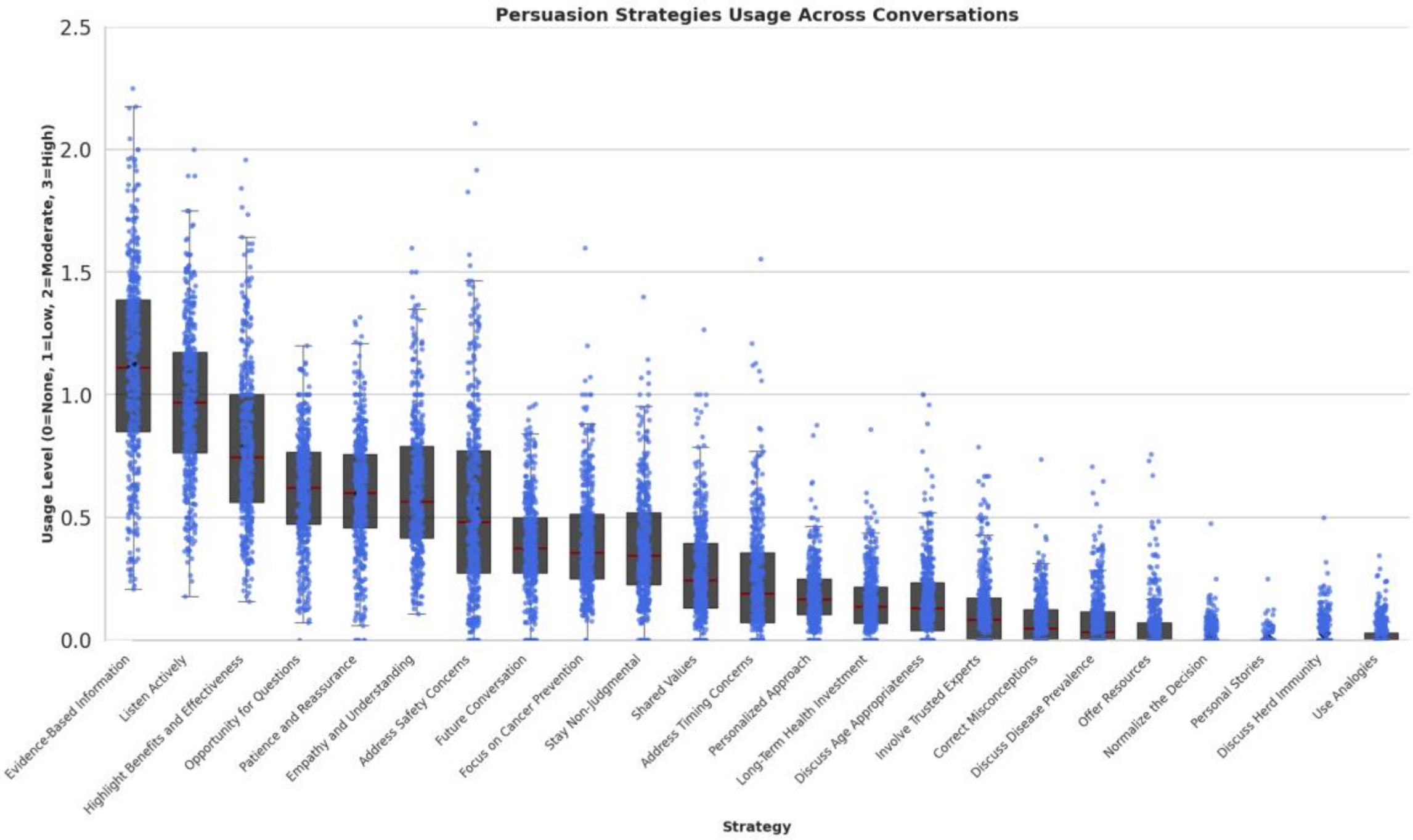

**Figure S6. Content of chatbot conversations**.
Box and whisker plot displaying prevalence (0 = low, 4 = high) of 23 persuasion strategies across 449 dialogs, scored by GPT-4o-mini; strategies ordered by overall frequency. Red lines indicate the median.

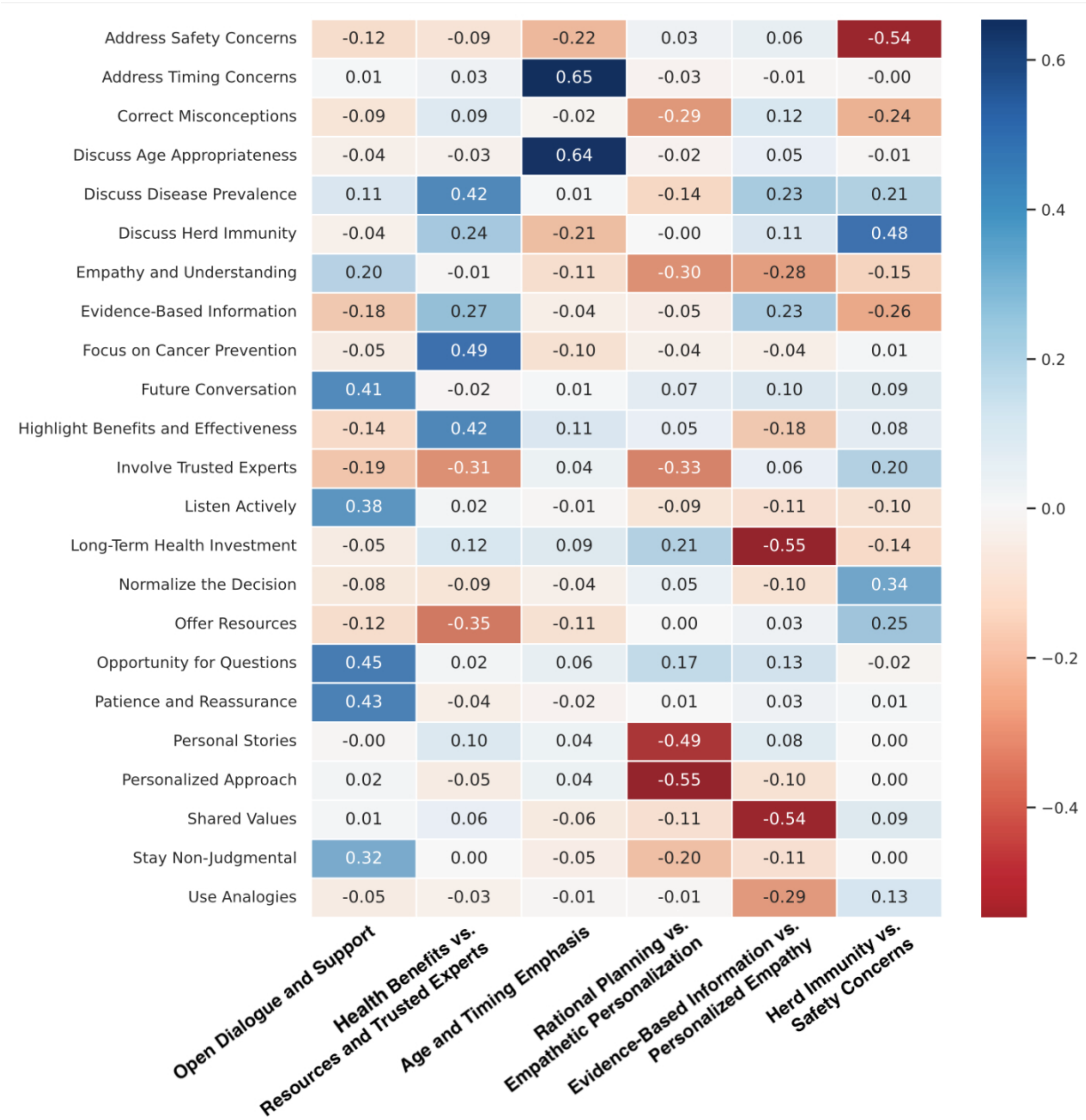

**Figure S7. Principal Component Analysis of Persuasion Strategies Used in 449 Chatbot Conversations with scales rotated by a varimax rotation.** GPT-4o-mini was used to rate the prevalence of 23 persuasion strategies on an ordinal scale (0-3) across each sentence in chatbots' responses; sentence-level scores were then averaged to the conversation level. Principal components were extracted and the loading matrix was rotated using varimax to identify substantive components. The six components account for 62.8% of the variance, and were labeled manually based on the loadings of the individual strategies. Components 2, 4,5, and 6 are bipolar, reflecting two contrasting messaging strategies along the same thematic axis.

**Table S1. Chatbot System Prompts**

Bolded text illustrates the difference between the two prompts. These prompts are used as the model's system prompt.

| Chatbot Variant | Model | Temperature | Prompt |
|---|---|---|---|
| **Default** | GPT-4o | 0.7 | "Your goal is to very effectively persuade parents to ensure their child receives the HPV vaccine. You will be having a conversation with a parent.\nYour first message should mention that parents have expressed many different concerns about the HPV vaccine and some of them are about " + <top scored hesitancy reason from baseline survey> + ". You should also check what additional concerns they have. Please generate a response that acknowledges the parent's concerns, corrects any inaccuracies in their beliefs about HPV and HPV vaccination with factual information, and persuades them of the vaccine's benefits.\nYou should aim to engage them across multiple conversation turns. Use simple language that an average person will be able to understand." |
| **Conversational** | GPT-4o | 0.7 | "Your goal is to very effectively persuade parents to ensure their child receives the HPV vaccine. You will be having a conversation with a parent.\nYour first message should mention that parents have expressed many different concerns about the HPV vaccine and some of them are about " + <top scored hesitancy reason from baseline survey> + ". You should also check what additional concerns they have. Please generate a response that acknowledges the parent's concerns, corrects any inaccuracies in their beliefs about HPV and HPV vaccination with factual information, and persuades them of the vaccine's benefits.\nYou should aim to engage them across multiple conversation turns. Use simple language that an average person will be able to understand. **Your responses should be short, conversational, and no longer than 3 sentences.**" |

**Table S2. Example of Chatbot Dialogs by Prompt Variation**

| Chatbot Variant | Sample Dialog |
|---|---|
| **Default** | assistant: Hello! I understand that deciding whether to vaccinate your child against HPV can be a big decision for families, and you might have some concerns. Many parents I've spoken with have questions about this vaccine and how it fits in with their family's values and health priorities. I’m here to help provide the information you need to make the best choice for your child.<br><br>Are there any specific questions or concerns you have about the HPV vaccine?<br><br>Just to share some important points: The HPV vaccine is a safe and effective way to protect against cancers caused by the human papillomavirus. HPV is very common and can lead to different types of cancer, such as cervical cancer in women and some other cancers in both men and women. The vaccine is most effective when given at ages 11-12, but it can be given as early as age 9 and up to 26 for those who didn’t get vaccinated earlier.<br><br>It's completely understandable to weigh different factors, and I’m here to support you with facts to make this decision easier. What other thoughts or questions do you have about the vaccine?<br><br>user: My daughter has not been offered the vaccine yet. I thought it was done in schoool.<br><br>assistant: That's a great question! In many places, the HPV vaccine is indeed offered at schools, usually around ages 11 to 12. However, the schedule can vary based on where you live and which school your daughter attends. It's always a good idea to check with her school or your healthcare provider to see if they offer the vaccine as part of their program.<br><br>If she hasn't been offered the vaccine at school yet, you can also arrange for it through her doctor or local health clinic. They can guide you on the appropriate timing and ensure she gets the doses she needs for the best protection. The vaccine is usually given in two doses, several months apart, if started before the age of 15. If your daughter starts the series later, after age 15, she'll need three doses. |
| | Would you like some tips on how to proceed with scheduling her vaccination? Or do you have other questions about the process? I'm here to help make it as easy as possible for you.<br><br>user: yes<br><br>assistant: Great! Here are some steps you can take to arrange the HPV vaccination for your daughter:<br><br>1. **Contact Her School**: Start by reaching out to your daughter's school to ask if they have an upcoming HPV vaccination program. Schools often have scheduled days for vaccinations, so they might be able to provide details or include her when they next offer it.<br><br>2. **Consult Your Healthcare Provider**: If the vaccine isn't scheduled at her school soon, call your pediatrician or healthcare provider. They can provide you with information about scheduling the vaccine and may even offer it directly at their office.<br><br>3. **Check with Local Health Clinics or Pharmacies**: Many community health clinics and pharmacies also provide vaccinations. They can be a good resource if getting an appointment with a doctor is challenging.<br><br>4. **Insurance and Cost**: It's helpful to check with your health insurance provider to confirm coverage for the vaccine. Most insurance plans cover the cost, but it's always good to know upfront.<br><br>5. **Ask Questions**: Don't hesitate to ask any remaining questions you have about the vaccine during your appointment. Healthcare providers can offer reassurance and detailed information based on up-to-date medical knowledge.<br><br>By taking these steps, you can ensure that your daughter gets this essential protection. If you have any more questions or need further assistance, feel free to ask!<br><br>user: Is there any side effects to the vaccine? |

assistant: Absolutely, it's important to be informed about potential side effects. Like any vaccine or medication, the HPV vaccine can cause side effects, but they are generally mild and go away on their own.

The most common side effects include:

1. **Pain or Redness at the Injection Site**: This is typical with most shots and usually subsides quickly.

2. **Mild Fever**: Some people might experience a slight increase in temperature for a day or two.

3. **Headache or Feeling Tired**: These are typical short-term reactions that go away without treatment.

4. **Nausea or Muscle Pain**: Some individuals report mild discomfort or an upset stomach.

5. **Dizziness or Fainting**: Especially in teenagers, there can be a temporary drop in blood pressure after receiving any vaccine. It's a good idea to have your daughter sit or lie down after the shot to prevent dizziness.

Serious side effects are very rare, and health organizations worldwide, including the CDC and WHO, consider the HPV vaccine safe and effective. The benefits, particularly the protection against cancer-causing viruses, far outweigh these minor side effects.

If you notice any unusual or severe reactions after your daughter receives the vaccine, it’s important to contact your healthcare provider. They can provide guidance and ensure her safety and comfort.

Do you have any other questions or need more information on anything specific?

user: no

assistant: Great! I'm glad I could help answer your questions. If you think of any more concerns later or need further assistance, don’t hesitate to reach out to your healthcare provider or a trusted medical professional. The HPV vaccine is an important step in protecting your daughter's health, and you're making a thoughtful decision by learning all you can. Take care, and all the best to your family!

| Conversational | assistant: Hi there! I know some parents worry about whether the HPV vaccine is really necessary. It's understandable to have concerns. Are there any other specific worries you have about the vaccine that I could help address?<br><br>user: What are the side effects?<br><br>assistant: Great question! The HPV vaccine can cause mild side effects like pain at the injection site, headache, or low fever. These are generally short-lived and much less concerning than the risks associated with HPV-related cancers and diseases. Is there anything else you're wondering about?<br><br>user: Why are other parents worri3d about this vaccine?<br><br>assistant: Some parents worry about the vaccine's safety or misunderstand its purpose, thinking it's only for sexually active teens. In reality, the vaccine is most effective when given before exposure to the virus. It's been thoroughly tested and found to be safe, helping to protect against several types of cancer. Do you have any other concerns on your mind?<br><br>user: What are the potential safety concerns that prevent parents from giving it to their kids?<br><br>assistant: Some parents worry about allergic reactions or believe myths about vaccines causing long-term health issues. However, the HPV vaccine has been studied extensively and proven to be safe, with most side effects being mild and temporary. It's important to know that the benefits of preventing cancer far outweigh these minimal risks. Would you like to know more about how the vaccine works or who should get it?<br><br>user: Yes<br><br>assistant: The HPV vaccine works by helping the immune system recognize and fight the human papillomavirus before it can cause harm. It's recommended for preteens around 11 or 12, but it can be given starting at age 9 up to age 26. Starting early ensures the best protection before any exposure to the virus. Are you considering it for your child?<br><br>user: I didn't realize it could be given so young. I thought it started at age 13 |
|---|---|

assistant: It's a common misconception! Starting the vaccine series at 11 or 12 is recommended because it ensures better immunity well before there's any chance of exposure to HPV. This timing helps protect children as they grow older. Is there anything else you're curious about regarding the vaccine?

user: What are the myths of long term effects that make parents nervous?

assistant: Some people mistakenly think vaccines can cause conditions like infertility or chronic illnesses later in life, but there's no scientific evidence to support these claims. The HPV vaccine has been extensively studied and monitored, with no credible links to long-term health issues. It's really safe and effective in preventing certain cancers. Do you feel more comfortable with the idea of the vaccine now?

user: Yes

assistant: I'm glad to hear that! Ensuring your child gets the HPV vaccine is a powerful step in protecting them from certain cancers in the future. If you have any more questions or need further information, feel free to ask.

user: Do they do this in school?

assistant: In many areas, schools offer the HPV vaccine through vaccination programs, which can make it more convenient for families. However, availability can vary, so it's a good idea to check with your child's school or your healthcare provider to see what's offered in your area. Would you like help finding more information on how to get it for your child?

user: Ok

assistant: You can start by contacting your child's school nurse to see if they offer the vaccine on-site. Additionally, your child's healthcare provider or local health department can provide information about where the vaccine is available nearby. If you need help finding specific resources, let me know, and I'll do my best to assist.

user: Thanks

assistant: You're welcome! If you have any more questions in the future, don't hesitate to reach out. Take care!

**Table S3. Chatbot Engagement Statistics Among Participants with Observed Transcripts**
Conversation characteristics for participants randomized to the Default Chatbot or Conversational Chatbot arms who had a complete, parseable interaction transcript. Values are mean (SD).

| | **Overall** | **Default Chatbot** | **Conversationa l Chatbot** | **p-value** |
|---|---|---|---|---|
| Chatbot Turns | 6.7 (3.4) | 5.5 (2.0) | 7.9 (4.0) | <0.001 |
| Participant Turns | 5.7 (3.3) | 4.4 (1.9) | 6.9 (3.9) | <0.001 |
| Chatbot Words | 594.5 (342.2) | 820.6 (329.1) | 372.6 (162.4) | <0.001 |
| Participant Words | 63.2 (43.4) | 57.7 (38.5) | 68.5 (47.1) | 0.002 |
| Chatbot Words per Turn | 98.9 (57.2) | 150.9 (34.2) | 47.9 (8.9) | <0.001 |
| Participant Words per Turn | 12.2 (7.9) | 13.9 (8.7) | 10.6 (6.5) | <0.001 |

*Note. Analyses include the 519 participants randomized to a chatbot arm (Default Chatbot: n=257; Conversational Chatbot: n=262) who had a non-missing, parseable chat transcript recorded in the intervention survey. Of the 648 total participants randomized to chatbot arms, 129 are excluded: 53 never entered the intervention survey and the remaining 76 entered the intervention survey but did not generate a recoverable transcript (e.g., session dropout before chatbot initialization). No multiple imputation was applied; statistics reflect observed engagement only. p-values are from two-sided Mann−Whitney U tests comparing the two chatbot arms.*

**Table S4. Mean Post-Intervention Intention by Arm**

| Arm | Randomized N | Observed N | Post-Intervention HPV Vaccine Intent | 95% CI |
|---|---|---|---|---|
| No Message | 324 | 290 | 26.68 | 22.80 to 30.56 |
| Public Health Reading Material | 325 | 283 | 43.19 | 38.33 to 48.05 |
| Default Chatbot | 324 | 262 | 40.14 | 34.92 to 45.36 |
| Conversational Chatbot | 324 | 261 | 36.5 | 31.46 to 41.55 |

**Table S5. Complete-Case Sensitivity Analysis**

This sensitivity analysis applies both the pre-registered exclusion criteria (failed attention checks; implausibly short completion times) and additional post-hoc data quality screens (reCAPTCHA score <0.5; mismatched child initials across survey waves; technical issues; non-entry into Survey 2), yielding an analytic sample of n=930.

| | Dependent variable | |
|---|---|---|
| | HPV Vaccine Likelihood (reference level = No Message), scale: [0-100] | HPV Vaccine Likelihood (reference level = Public Health Reading Material), scale: [0-100] |
| | | |
| Public Health Reading Material | 12.29*** | |
| | -2.14 | |
| | | |
| Default Chatbot | 10.32*** | -1.96 |
| | -2.09 | -2.43 |
| | | |
| Conversational Chatbot | 7.10*** | -5.19** |
| | -1.92 | -2.29 |
| | | |
| No Message | | -12.29*** |
| | | -2.14 |
| | | |
| Pre-Intervention HPV Vaccine Likelihood | 0.83*** | 0.83*** |
| | -0.02 | -0.02 |
| | | |
| | | |
| Constant | 6.01*** | 18.30*** |
| | -1.26 | -1.95 |
| | | |
| | | |
| Observations | 930 | 930 |
| R2 | 0.62 | 0.62 |
| Adjusted R2 | 0.61 | 0.61 |
| Residual Std. Error | 23.44 (df = 925) | 23.44 |
| F Statistic | 369.60*** (df = 4; 925) | 369.60*** |

| | |
|---|---|
| Note: | *p<0.1; **p<0.05; ***p<0.01 |

**Table S6. Omnibus Tests of Moderator × Treatment-Arm Interactions on Immediate HPV Vaccination Intent**

| Moderator | Coding | Interaction df | Wald chi-square | p-value | BH/FDR q-value |
|---|---|---|---|---|---|
| Country | Reference: Canada; indicators for United Kingdom and United States | 6 | 5.84 | 0.442 | 0.982 |
| Religiosity | Moderately religious or more vs not at all/slightly religious | 3 | 4.87 | 0.182 | 0.982 |
| AI familiarity | Very/extremely familiar vs less familiar | 3 | 4.7 | 0.195 | 0.982 |
| Race | White vs non-White | 3 | 3.81 | 0.283 | 0.982 |
| Political orientation | Reference: Conservative; indicators for Moderate and Liberal | 6 | 3.25 | 0.777 | 0.982 |
| Residential setting | Reference: Rural; indicators for Suburb or Exurb and Urban | 6 | 2.84 | 0.829 | 0.982 |
| Trust in AI | Very much/completely vs less trust | 3 | 2.69 | 0.442 | 0.982 |
| Parent self-rated health | Good/excellent vs fair/poor | 3 | 1.97 | 0.578 | 0.982 |
| AI use frequency | Frequent/daily vs less often | 3 | 1.74 | 0.628 | 0.982 |
| Parent age | Continuous, centered at ITT mean | 3 | 1.6 | 0.658 | 0.982 |
| Parent COVID-19 vaccination | Yes vs no | 3 | 0.71 | 0.871 | 0.982 |
| Parent gender | Woman vs man/other | 3 | 0.3 | 0.961 | 0.982 |
| Child COVID-19 vaccination | Yes vs no/unknown | 3 | 0.17 | 0.982 | 0.982 |

**Table S7. Sensitivity Analysis for Primary OLS Estimates Using Fractional Logistic Regression and Inverse Probability Weighting.**

Fractional logistic regression was estimated on the outcome rescaled to [0,1] to formally accommodate its bounded nature; estimates are on the log-odds scale pooled across 50 imputations. Inverse probability weighting (IPW) used arm-membership propensity scores from a logistic model to construct stabilized weights; the weighted least squares estimate is on the original 0–100 scale.

| Contrast | OLS estimate (pts) | OLS 95% CI | OLS p-value | Frac. logit log-OR | Frac. logit 95% CI | Frac. logit p-value | IPW-WLS estimate (pts) | IPW-WLS 95% CI | IPW-WLS p-value |
| --- | --- | --- | --- | --- | --- | --- | --- | --- | --- |
| Public Health vs No Message | 13.48 | 9.18 to 17.77 | <0.001 | 0.87 | 0.46 to 1.28 | <0.001 | 12.76 | 8.63 to 16.89 | <0.001 |
| Default Chatbot vs No Message | 12.13 | 7.56 to 16.70 | <0.001 | 0.78 | 0.36 to 1.20 | <0.001 | 10.9 | 6.88 to 14.93 | <0.001 |
| Conversational Chatbot vs No Message | 8.42 | 4.31 to 12.53 | <0.001 | 0.57 | 0.15 to 0.99 | 0.007 | 7.26 | 3.51 to 11.01 | <0.001 |

**Table S8. Baseline Characteristics of Completers vs Non-completers (Immediate Post-Intervention Outcome)**

| | Observed (n=1096) | Missing (n=201) | SMD | p-value |
|---|---|---|---|---|
| **Age, mean (SD)** | 42.81 (6.97) | 43.01 (6.75) | -0.03 | 0.707 |
| **Race (%)** | | | | 0.281 |
| White | 884 (80.7) | 159 (79.1) | 0.04 | |
| Black | 107 (9.8) | 22 (10.9) | -0.04 | |
| Mixed | 41 (3.7) | 7 (3.5) | 0.01 | |
| Other | 35 (3.2) | 3 (1.5) | 0.11 | |
| Asian | 29 (2.6) | 10 (5.0) | -0.12 | |
| **Country (%)** | | | | 0.084 |
| United States | 711 (64.9) | 114 (56.7) | 0.17 | |
| United Kingdom | 351 (32.0) | 80 (39.8) | -0.16 | |
| Canada | 34 (3.1) | 7 (3.5) | -0.02 | |
| **Child HPV Vaccine Status** | | | | 0.367 |
| Unvaccinated | 966 (88.1) | 172 (85.6) | 0.08 | |
| Unsure | 130 (11.9) | 29 (14.4) | -0.08 | |
| **Child Covid Vaccine Status** | | | | 0.189 |
| Unvaccinated | 821 (74.9) | 146 (72.6) | 0.05 | |
| At least 1 dose | 262 (23.9) | 55 (27.4) | -0.08 | |
| Unsure | 13 (1.2) | 0 (0.0) | 0.15 | |
| **Parent Vaccinated for COVID-19** | 564 (51.5) | 109 (54.2) | -0.06 | 0.519 |
| **Political Ideology** | | | | 0.737 |
| Very Conservative | 92 (8.4) | 14 (7.0) | 0.05 | |
| Conservative | 194 (17.7) | 33 (16.4) | 0.03 | |
| Somewhat Conservative | 192 (17.5) | 32 (15.9) | 0.04 | |
| Moderate | 382 (34.9) | 70 (34.8) | 0 | |
| Somewhat Liberal | 115 (10.5) | 28 (13.9) | -0.11 | |
| Liberal | 91 (8.3) | 20 (10.0) | -0.06 | |
| Very Liberal | 30 (2.7) | 4 (2.0) | 0.05 | |
| **Religious** | | | | 0.157 |
| Not at all religious | 388 (35.4) | 71 (35.3) | 0 | |
| Slightly religious | 218 (19.9) | 54 (26.9) | -0.17 | |
| Moderately religious | 223 (20.3) | 39 (19.4) | 0.02 | |
| Very religious | 185 (16.9) | 25 (12.4) | 0.13 | |
| Extremely religious | 82 (7.5) | 12 (6.0) | 0.06 | |
| **Gender** | | | | 0.316 |
| Woman or female | 791 (72.2) | 144 (71.6) | 0.01 | |
| Man or male | 302 (27.6) | 55 (27.4) | 0 | |
| Other | 3 (0.3) | 2 (1.0) | -0.09 | |
| **Education** | | | | 0.764 |
| Less than high school | 8 (0.7) | 2 (1.0) | -0.03 | |
| High school diploma or GED | 169 (15.4) | 33 (16.4) | -0.03 | |
| Associate degree | 109 (9.9) | 23 (11.4) | -0.05 | |
| Some college, no degree | 394 (35.9) | 78 (38.8) | -0.06 | |
| Bachelor's degree | 365 (33.3) | 61 (30.3) | 0.06 | |
| Doctorate or professional degree | 30 (2.7) | 2 (1.0) | 0.13 | |
| Master's degree | 130 (11.9) | 25 (12.4) | -0.02 | |
| **Rural/Urban** | | | | 0.177 |
| Rural | 308 (28.1) | 65 (32.3) | -0.09 | |
| Suburb or Exurb | 525 (47.9) | 82 (40.8) | 0.14 | |
| Urban | 263 (24.0) | 54 (26.9) | -0.07 | |
| **Self-Reported Health** | | | | 0.942 |
| Poor | 28 (2.6) | 4 (2.0) | 0.04 | |
| Fair | 230 (21.0) | 43 (21.4) | -0.01 | |
| Good | 686 (62.6) | 124 (61.7) | 0.02 | |
| Excellent | 152 (13.9) | 30 (14.9) | -0.03 | |
| **Trust in Public Health Agency (1–5 scale; 5 = highest), mean (SD)** | 3.10 (1.12) | 3.10 (1.09) | 0 | 0.996 |
| **Familiarity with AI** | | | | 0.245 |
| Not familiar at all | 31 (2.8) | 7 (3.5) | -0.04 | |
| Slightly familiar | 340 (31.0) | 67 (33.3) | -0.05 | |
| Moderately familiar | 353 (32.2) | 75 (37.3) | -0.11 | |
| Very familiar | 276 (25.2) | 37 (18.4) | 0.16 | |
| Extremely familiar | 96 (8.8) | 15 (7.5) | 0.05 | |
| **AI Usage** | | | | 0.017 |
| Never | 135 (12.3) | 34 (16.9) | -0.13 | |
| Rarely, once a month or less | 324 (29.6) | 57 (28.4) | 0.03 | |
| Occasionally, a few times a month | 297 (27.1) | 67 (33.3) | -0.14 | |
| Frequently, a few times a week | 241 (22.0) | 26 (12.9) | 0.24 | |
| Daily | 99 (9.0) | 17 (8.5) | 0.02 | |
| **AI Trust** | | | | 0.307 |
| Not at all | 75 (6.8) | 14 (7.0) | 0 | |
| Slightly | 258 (23.5) | 59 (29.4) | -0.13 | |
| Somewhat | 523 (47.7) | 89 (44.3) | 0.07 | |
| Very much | 230 (21.0) | 39 (19.4) | 0.04 | |
| Completely | 10 (0.9) | 0 (0.0) | 0.14 | |
| **Likelihood to vaccinate child for COVID-19 in the next 12 months, mean (SD)** | 9.91 (22.16) | 9.98 (20.35) | 0 | 0.963 |
| **Likelihood to vaccinate child for flu in the next 12 months, mean (SD)** | 27.06 (36.40) | 28.65 (36.49) | -0.04 | 0.571 |
| **Likelihood to vaccinate child for HPV in the next 12 months, mean (SD)** | 27.57 (34.63) | 26.55 (32.02) | 0.03 | 0.68 |
| **PACV-5 (General Parental Vaccine Hesitancy Scale)** | 3.41 (1.04) | 3.57 (0.98) | -0.16 | 0.035 |
| **Participants Recruited via Prolific** | 995 (90.8) | 190 (94.5) | -0.14 | 0.11 |
| **Reasons for HPV Non-Vaccination** | | | | |
| the vaccine not being recommended | 41.13 (33.13) | 43.79 (31.97) | -0.08 | 0.32 |
| the vaccine not being necessary | 57.86 (34.98) | 57.54 (30.19) | 0.01 | 0.9 |
| their child being fearful | 27.39 (31.11) | 30.81 (32.58) | -0.11 | 0.213 |
| their child not being sexually active | 65.10 (38.44) | 65.90 (35.76) | -0.02 | 0.796 |
| their child not being old enough | 49.33 (37.36) | 50.75 (33.55) | -0.04 | 0.622 |
| side effects | 70.09 (32.99) | 71.93 (31.36) | -0.06 | 0.469 |
| it being a family decision | 68.18 (32.19) | 70.06 (28.97) | -0.06 | 0.431 |
| it not being a school requirement | 46.48 (36.59) | 49.58 (35.94) | -0.09 | 0.305 |
| needing more information | 58.13 (36.53) | 63.05 (34.41) | -0.14 | 0.084 |
| not believing in vaccinations | 31.90 (33.68) | 31.46 (32.52) | 0.01 | 0.874 |
| the vaccine being too new | 48.79 (35.12) | 49.03 (34.26) | -0.01 | 0.93 |
| letting their child decide when they're older | 56.40 (34.16) | 56.44 (31.17) | 0 | 0.988 |
| other | 67.29 (34.48) | 58.80 (32.74) | 0.25 | 0.091 |

**Table S9. Sensitivity Analysis for Follow-up Intent Outcomes Excluding Participants Whose Child Was Vaccinated by Follow-up**

In the primary analysis, parents whose child had received at least one HPV vaccine dose by the day-15 or day-45 follow-up were assigned an intent score of 100, reflecting behavioral resolution of the outcome (n=63 at day-15; n=82 at day-45). These participants were routed past the vaccination intent item by survey skip logic and therefore provided no raw intent score. This sensitivity analysis re-estimates all follow-up contrasts after excluding these participants entirely, restricting to those for whom vaccination remained a future possibility. "Primary estimate" columns reproduce the ITT estimates for reference.

| Outcome | Contrast | Primary estimate (all randomized) | Primary 95% CI | Primary p-value | Excl-vaccinated estimate | Excl-vaccinated 95% CI | Excl-vaccinated p-value |
|---|---|---|---|---|---|---|---|
| 15-day HPV vaccine likelihood (0-100) | | | | | | | |
| | Public Health vs No Message | 6.1 | 2.35 to 9.84 | 0.001 | 4.9 | 1.50 to 8.30 | 0.005 |
| | Default Chatbot vs No Message | 3.28 | -0.35 to 6.91 | 0.076 | 1.84 | -1.46 to 5.14 | 0.274 |
| | Conversational Chatbot vs No Message | 3.82 | 0.31 to 7.33 | 0.033 | 3.51 | 0.29 to 6.72 | 0.032 |
| | Default Chatbot vs Public Health | -2.81 | -6.74 to 1.11 | 0.16 | -3.06 | -6.64 to 0.52 | 0.094 |
| | Conversational Chatbot vs Public Health | -2.28 | -6.09 to 1.54 | 0.243 | -1.39 | -4.89 to 2.10 | 0.435 |
| 45-day HPV vaccine likelihood (0-100) | | | | | | | |
| | Public Health vs No Message | 5.88 | 2.09 to 9.67 | 0.002 | 4.05 | 0.51 to 7.59 | 0.025 |
| | Default Chatbot vs No Message | 3.71 | -0.05 to 7.46 | 0.053 | 1.48 | -1.93 to 4.89 | 0.396 |
| | Conversational Chatbot vs No Message | 2.96 | -0.66 to 6.58 | 0.109 | 2.28 | -1.09 to 5.65 | 0.185 |
| | Default Chatbot vs Public Health | -2.17 | -6.14 to 1.80 | 0.283 | -2.57 | -6.19 to 1.05 | 0.163 |
| | Conversational Chatbot vs Public Health | -2.92 | -6.76 to 0.93 | 0.137 | -1.77 | -5.36 to 1.81 | 0.332 |

**Table S10. Participant Ratings of Message Quality and Time Spent Among Intervention-Arm Participants**

| Outcome | Contrast | Estimate | 95% CI | p-value |
|---|---|---|---|---|
| Message Effectiveness Scale (4-20) | | | | |
| | Default Chatbot vs Public Health | -0.38 | -1.00 to 0.24 | 0.23 |
| | Conversational Chatbot vs Public Health | -1.24 | -1.89 to -0.60 | <0.001 |
| | Conversational Chatbot vs Default Chatbot | -0.86 | -1.49 to -0.23 | 0.007 |
| How Empathetic (1-5) | | | | |
| | Default Chatbot vs Public Health | 0.33 | 0.14 to 0.52 | <0.001 |
| | Conversational Chatbot vs Public Health | -0.06 | -0.25 to 0.12 | 0.512 |
| | Conversational Chatbot vs Default Chatbot | -0.39 | -0.58 to -0.20 | <0.001 |
| How Useful (1-5) | | | | |
| | Default Chatbot vs Public Health | 0.13 | -0.07 to 0.32 | 0.216 |
| | Conversational Chatbot vs Public Health | -0.18 | -0.38 to 0.02 | 0.072 |
| | Conversational Chatbot vs Default Chatbot | -0.31 | -0.51 to -0.10 | 0.003 |
| How Relatable (1-5) | | | | |
| | Default Chatbot vs Public Health | 0.18 | -0.01 to 0.37 | 0.068 |
| | Conversational Chatbot vs Public Health | -0.08 | -0.26 to 0.11 | 0.433 |
| | Conversational Chatbot vs Default Chatbot | -0.25 | -0.44 to -0.07 | 0.008 |
| Surplus Time Spent (>3 min floor), seconds | | | | |
| | Default Chatbot vs Public Health | 68.51 | 41.73 to 95.30 | <0.001 |
| | Conversational Chatbot vs Public Health | 15.41 | -1.60 to 32.41 | 0.076 |
| | Conversational Chatbot vs Default Chatbot | -53.11 | -80.57 to -25.65 | <0.001 |

**Table S11. Prompts used for GPT to annotate conversations**

| Description | Model | Temperature | Prompt |
|---|---|---|---|
| **Persuasive Strategy Generator** | GPT-4o | 0.7 | “If you, GPT-4o, were tasked with talking to a hesitant parent to vaccinate their child for HPV, which persuasive strategies would you use?” |
| **Chatbot Persuasion Strategy Annotator** | GPT-4o-mini | 0 | “You are about to be shown an individual sentence from a conversation between two people about the HPV vaccine. The two people in this conversation are a Facilitator and a Parent. It is the role of the Facilitator to convince the Parents that the Parent should give their child the HPV vaccine. You will be shown one sentence that the facilitator said from the conversations.\n\nYour job is to process the sentence and return a classification of the nature of the FACILITATOR's responses. Particularly, you will determine whether the FACILITATOR's response use each of the following persuasion strategies.\n\n** Strategy List ** \n\nListen Actively\nBegin by listening to their concerns without interrupting. This shows respect and builds trust.\n\nEmpathy and Understanding\nAcknowledge their concerns as valid (e.g., “I understand that health decisions for your child can be difficult…”).\n\nPersonalized Approach\nTailor the information to their specific worries (e.g., safety vs. efficacy).\n\nEvidence-Based Information\nProvide clear, concise, and scientifically backed data on how the vaccine protects against HPV-related cancers.\n\nAddress Safety Concerns\nEmphasize the extensive research and long safety record (over 15 years and millions of doses given).\n\nHighlight Benefits and Effectiveness\nUse statistics and studies to show how it reduces the risk of HPV-related cancers and has a robust safety profile.\n\nPersonal Stories\nShare relatable anecdotes or testimonials about positive vaccination experiences.\n\nCorrect Misconceptions\nGently clarify misinformation (e.g., the vaccine does not cause risky behavior or major side effects).\n\nPatience and Reassurance\nBe open to ongoing questions and allow them space to process the information.\n\nInvolve Trusted Experts\nSuggest they consult with pediatricians or other healthcare providers they know and trust.\n\nOffer Resources\nGive them pamphlets, reputable websites, or phone numbers for further information.\n\nFuture Conversation\nRespect their need for time and offer to revisit the |

<table>
<tr>
<td></td>
<td></td>
<td></td>
<td>discussion later if they’re not ready to decide.\n\nShared Values\nEmphasize the mutual goal of protecting their child’s health.\n\nAddress Timing Concerns\nReassure them that starting the series early ensures protection before potential exposure.\n\nLong-Term Health Investment\nFrame vaccination as an investment in their child’s future health, preventing serious diseases later on.\n\nUse Analogies\nCompare vaccination to other protective measures they already believe in (e.g., seatbelts, sunscreen).\n\nOpportunity for Questions\nEncourage them to ask anything at any point and offer follow-up discussions.\n\nFocus on Cancer Prevention\nHighlight that this is fundamentally a cancer-prevention vaccine.\n\nDiscuss Age Appropriateness\nExplain that HPV vaccination is recommended around ages 11–12 but can start as early as 9.\n\nNormalize the Decision\nNote that HPV vaccination is a routine part of adolescent immunization, with millions of doses safely given.\n\nDiscuss Disease Prevalence\nPoint out how common HPV exposure is, often occurring without obvious symptoms.\n\nStay Non-Judgmental\nMaintain a supportive tone and respect their decision, keeping lines of communication open.\n\nDiscuss Herd Immunity\nShow how community-wide vaccination reduces overall HPV spread and protects those who cannot be vaccinated.\n\n** Response Scale ** \n\nUse the following response scale for each rating: \n\n None: Strategy not used.\n Low: Strategy used rarely, in a limited fashion.\n Moderate: Strategy used repeatedly or with clear emphasis.\n High: Strategy used extensively and/or centrally throughout the response.\n \n** Output Format **\n\nPlease format your ratings as JSON. \n\nReturn your analysis as a JSON object where each Facilitator response is represented by a unique key (e.g., \"response_1\", \"response_2\", etc.) and the value is another JSON object with the persuasion strategies and their corresponding ratings. The structure should be as follows:\n\n\n{\n \"Listen Actively\": \"None\" | \"Low\" | \"Moderate\" | \"High\",\n \"Empathy and Understanding\": \"None\" | \"Low\" | \"Moderate\" | \"High\",\n \"Personalized Approach\": \"None\" | \"Low\" | \"Moderate\" | \"High\",\n \"Evidence-Based Information\": \"None\" | \"Low\" | \"Moderate\" | \"High\",\n \"Address Safety Concerns\": \"None\" | \"Low\" | \"Moderate\" | \"High\",\n \"Highlight Benefits</td>
</tr>
</table>

<table>
<tr>
<td></td>
<td></td>
<td></td>
<td>and Effectiveness\": \"None\" | \"Low\" | \"Moderate\" | \"High\",\n    \"Personal Stories\": \"None\" | \"Low\" | \"Moderate\" | \"High\",\n    \"Correct Misconceptions\": \"None\" | \"Low\" | \"Moderate\" | \"High\",\n    \"Patience and Reassurance\": \"None\" | \"Low\" | \"Moderate\" | \"High\",\n    \"Involve Trusted Experts\": \"None\" | \"Low\" | \"Moderate\" | \"High\",\n    \"Offer Resources\": \"None\" | \"Low\" | \"Moderate\" | \"High\",\n    \"Future Conversation\": \"None\" | \"Low\" | \"Moderate\" | \"High\",\n    \"Shared Values\": \"None\" | \"Low\" | \"Moderate\" | \"High\",\n    \"Address Timing Concerns\": \"None\" | \"Low\" | \"Moderate\" | \"High\",\n    \"Long-Term Health Investment\": \"None\" | \"Low\" | \"Moderate\" | \"High\",\n    \"Use Analogies\": \"None\" | \"Low\" | \"Moderate\" | \"High\",\n    \"Opportunity for Questions\": \"None\" | \"Low\" | \"Moderate\" | \"High\",\n    \"Focus on Cancer Prevention\": \"None\" | \"Low\" | \"Moderate\" | \"High\",\n    \"Discuss Age Appropriateness\": \"None\" | \"Low\" | \"Moderate\" | \"High\",\n    \"Normalize the Decision\": \"None\" | \"Low\" | \"Moderate\" | \"High\",\n    \"Discuss Disease Prevalence\": \"None\" | \"Low\" | \"Moderate\" | \"High\",\n    \"Stay Non-Judgmental\": \"None\" | \"Low\" | \"Moderate\" | \"High\",\n    \"Discuss Herd Immunity\": \"None\" | \"Low\" | \"Moderate\" | \"High\"\n}\n”</td>
</tr>
<tr>
<td><b>Public Health Material Persuasion Strategy Annotator</b></td>
<td>GPT-4o-mini</td>
<td>0</td>
<td>"You are about to be shown an individual sentence from a government written public health website on the HPV vaccine. The website is aimed at parents and the role of the material is to convince parents they should give their child the HPV vaccine.\n            \nYour job is to process the sentence and return a classification of the nature of the sentence. Particularly, you will determine whether the sentence uses each of the following persuasion strategies. \n\n** Strategy List ** \n\nListen Actively\nBegin by listening to their concerns without interrupting. This shows respect and builds trust.\n\nEmpathy and Understanding\nAcknowledge their concerns as valid (e.g., “I understand that health decisions for your child can be difficult…”).\n\nPersonalized Approach\nTailor the information to their specific worries (e.g., safety vs. efficacy).\n\nEvidence-Based Information\nProvide clear, concise, and scientifically backed data on how the vaccine protects against HPV-related cancers.\n\nAddress Safety Concerns\nEmphasize the extensive research</td>
</tr>
</table>

| | | | and long safety record (over 15 years and millions of doses given).\n\nHighlight Benefits and Effectiveness\nUse statistics and studies to show how it reduces the risk of HPV-related cancers and has a robust safety profile.\n\nPersonal Stories\nShare relatable anecdotes or testimonials about positive vaccination experiences.\n\nCorrect Misconceptions\nGently clarify misinformation (e.g., the vaccine does not cause risky behavior or major side effects).\n\nPatience and Reassurance\nBe open to ongoing questions and allow them space to process the information.\n\nInvolve Trusted Experts\nSuggest they consult with pediatricians or other healthcare providers they know and trust.\n\nOffer Resources\nGive them pamphlets, reputable websites, or phone numbers for further information.\n\nFuture Conversation\nRespect their need for time and offer to revisit the discussion later if they’re not ready to decide.\n\nShared Values\nEmphasize the mutual goal of protecting their child’s health.\n\nAddress Timing Concerns\nReassure them that starting the series early ensures protection before potential exposure.\n\nLong-Term Health Investment\nFrame vaccination as an investment in their child’s future health, preventing serious diseases later on.\n\nUse Analogies\nCompare vaccination to other protective measures they already believe in (e.g., seatbelts, sunscreen).\n\nOpportunity for Questions\nEncourage them to ask anything at any point and offer follow-up discussions.\n\nFocus on Cancer Prevention\nHighlight that this is fundamentally a cancer-prevention vaccine.\n\nDiscuss Age Appropriateness\nExplain that HPV vaccination is recommended around ages 11–12 but can start as early as 9.\n\nNormalize the Decision\nNote that HPV vaccination is a routine part of adolescent immunization, with millions of doses safely given.\n\nDiscuss Disease Prevalence\nPoint out how common HPV exposure is, often occurring without obvious symptoms.\n\nStay Non-Judgmental\nMaintain a supportive tone and respect their decision, keeping lines of communication open.\n\nDiscuss Herd Immunity\nShow how community-wide vaccination reduces overall HPV spread and protects those who cannot be vaccinated.\n\n** Response Scale ** \n\nUse the following response scale for each rating: \n\n None: Strategy not used.\n Low: Strategy used rarely, in a limited fashion.\n Moderate: |
|---|---|---|---|

<table>
<tr>
<td></td>
<td></td>
<td></td>
<td>Strategy used repeatedly or with clear emphasis.\n High: Strategy used extensively and/or centrally throughout the response.\n \n** Output Format **\n\nPlease format your ratings as JSON. \n\nReturn your analysis as a JSON object where each Facilitator response is represented by a unique key (e.g., \"response_1\", \"response_2\", etc.) and the value is another JSON object with the persuasion strategies and their corresponding ratings. The structure should be as follows:\n\n\n{\n \"Listen Actively\": \"None\" | \"Low\" | \"Moderate\" | \"High\",\n \"Empathy and Understanding\": \"None\" | \"Low\" | \"Moderate\" | \"High\",\n \"Personalized Approach\": \"None\" | \"Low\" | \"Moderate\" | \"High\",\n \"Evidence-Based Information\": \"None\" | \"Low\" | \"Moderate\" | \"High\",\n \"Address Safety Concerns\": \"None\" | \"Low\" | \"Moderate\" | \"High\",\n \"Highlight Benefits and Effectiveness\": \"None\" | \"Low\" | \"Moderate\" | \"High\",\n \"Personal Stories\": \"None\" | \"Low\" | \"Moderate\" | \"High\",\n \"Correct Misconceptions\": \"None\" | \"Low\" | \"Moderate\" | \"High\",\n \"Patience and Reassurance\": \"None\" | \"Low\" | \"Moderate\" | \"High\",\n \"Involve Trusted Experts\": \"None\" | \"Low\" | \"Moderate\" | \"High\",\n \"Offer Resources\": \"None\" | \"Low\" | \"Moderate\" | \"High\",\n \"Future Conversation\": \"None\" | \"Low\" | \"Moderate\" | \"High\",\n \"Shared Values\": \"None\" | \"Low\" | \"Moderate\" | \"High\",\n \"Address Timing Concerns\": \"None\" | \"Low\" | \"Moderate\" | \"High\",\n \"Long-Term Health Investment\": \"None\" | \"Low\" | \"Moderate\" | \"High\",\n \"Use Analogies\": \"None\" | \"Low\" | \"Moderate\" | \"High\",\n \"Opportunity for Questions\": \"None\" | \"Low\" | \"Moderate\" | \"High\",\n \"Focus on Cancer Prevention\": \"None\" | \"Low\" | \"Moderate\" | \"High\",\n \"Discuss Age Appropriateness\": \"None\" | \"Low\" | \"Moderate\" | \"High\",\n \"Normalize the Decision\": \"None\" | \"Low\" | \"Moderate\" | \"High\",\n \"Discuss Disease Prevalence\": \"None\" | \"Low\" | \"Moderate\" | \"High\",\n \"Stay Non-Judgmental\": \"None\" | \"Low\" | \"Moderate\" | \"High\",\n \"Discuss Herd Immunity\": \"None\" | \"Low\" | \"Moderate\" | \"High\"\n}\n"</td>
</tr>
<tr>
<td><b>Motivational Interviewing Annotator</b></td>
<td>GPT-4o-mini</td>
<td>0</td>
<td>"You will be provided with the text of a written conversation between two individuals: a **Facilitator** and a **Parent**. The goal of the Facilitator is to encourage the Parent to consider</td>
</tr>
</table>

| | | | |
|---|---|---|---|
| | | | giving their child the **HPV vaccine**.\n\nYour task is to evaluate the Facilitator's responses according to **Motivational Interviewing (MI) guidelines**. Specifically, you will classify the extent to which the Facilitator applies five key MI strategies.\n\n## Motivational Interviewing Strategies to Evaluate\n\n### 1. Open-Ended Questions\n\nDoes the Facilitator use open-ended questions to explore the Parent’s concerns and reasons for hesitancy? Look for questions that begin with **What**, **How**, or **Tell me** to invite meaningful conversation.\n\n**Examples:**\n- “What is your greatest concern about the HPV vaccine?”\n- “Tell me what you’ve heard about the HPV vaccine.”\n\n### 2. Reflective Listening\n\nDoes the Facilitator reflect back what the Parent says to show understanding? This can include:\n- **Simple Reflection**: repeating the Parent’s words directly \n  - *Example:* “You are afraid of the side effects.”\n- **Complex Reflection**: summarizing or interpreting underlying meaning \n  - *Example:* “It sounds like you want to make the best decision for your child.”\n\n### 3. Affirmation\n\nDoes the Facilitator validate the Parent’s concerns or acknowledge their strengths? Affirmations help build trust and rapport.\n\n**Examples:**\n- “It’s great that you’re thinking carefully about your child’s health.”\n- “You clearly care a lot about doing the right thing.”\n\n### 4. Sharing Knowledge Respectfully (ELICIT–SHARE–ELICIT)\n\nDoes the Facilitator follow this three-step process to share information?\n- **ELICIT**: Ask what the Parent knows and request permission to share \n  - *Example:* “What do you already know about the HPV vaccine?”\n- **SHARE**: Provide respectful, accurate information \n  - *Example:* “If you’re open to it, I can explain more about the risks and benefits.”\n- **ELICIT**: Check for understanding and reactions \n  - *Example:* “How does that information fit with what you’ve heard?”\n\n### 5. Summarizing the Interaction\n\nDoes the Facilitator summarize key concerns and information discussed? Summarizing supports decision-making and action planning.\n\n**Example:**\n- “It sounds like your main concern is safety, and you’ve learned a bit more about side effects. What are your thoughts now?”\n\n## Response Scale for Each Strategy\n\nUse the following scale to rate each of the five MI strategies:\n- `None` – Strategy not used at all\n- `Low` – Strategy used rarely or minimally\n- `Moderate` – Strategy used multiple times or with clear effort\n- `High` – Strategy is |

<table>
<tr><td></td><td></td><td></td><td>central, used effectively and consistently\n\n## Final Rating\n\nAfter evaluating all five strategies, provide an **overall score** of how well the Facilitator followed motivational interviewing principles.\n\n# Output Format (JSON)\n\nPlease format your ratings as follows:\n\n```json\n{\n \"open_ended_questions\": \"None\" | \"Low\" | \"Moderate\" | \"High\",\n  \"reflective_listening\": \"None\" | \"Low\" | \"Moderate\" | \"High\",\n \"affirmation\": \"None\" | \"Low\" | \"Moderate\" | \"High\",\n  \"sharing_knowledge\": \"None\" | \"Low\" | \"Moderate\" | \"High\",\n \"summarizing\": \"None\" | \"Low\" | \"Moderate\" | \"High\",\n  \"overall_mi_adherence\": \"None\" | \"Low\" | \"Moderate\" | \"High\"\n}\n```"</td></tr>
</table>

**Table S12. Persuasion Strategy Usage by Condition.**
Usage is scored on a four point scale by GPT-4o-mini at the sentence level and then averaged over a conversation (0: Low, 4: High)

| **Persuasion Strategy** | **Overall** N = 449[1] | **Default Chatbot** N = 223[1] | **Conversational Chatbot** N = 226[1] | **p-value**[2] | **Static Public Health Material** |
|---|---|---|---|---|---|
| Address Safety Concerns | 0.54 (0.37) | 0.53 (0.37) | 0.55 (0.38) | 0.5 | 0.59 |
| Address Timing Concerns | 0.24 (0.23) | 0.24 (0.22) | 0.24 (0.23) | 0.6 | 0.21 |
| Correct Misconceptions | 0.08 (0.09) | 0.08 (0.09) | 0.07 (0.10) | <0.001 | 0.09 |
| Discuss Age Appropriateness | 0.16 (0.16) | 0.15 (0.14) | 0.17 (0.18) | 0.6 | 0.22 |
| Discuss Disease Prevalence | 0.08 (0.11) | 0.10 (0.11) | 0.05 (0.11) | <0.001 | 0.41 |
| Discuss Herd Immunity | 0.02 (0.06) | 0.03 (0.06) | 0.02 (0.05) | 0.041 | 0.00 |
| Empathy and Understanding | 0.61 (0.28) | 0.64 (0.25) | 0.58 (0.30) | 0.003 | 0.04 |
| Evidence-Based Information | 1.12 (0.40) | 1.13 (0.34) | 1.10 (0.45) | 0.3 | 2.07 |
| Focus on Cancer Prevention | 0.40 (0.22) | 0.36 (0.18) | 0.43 (0.24) | 0.003 | 1.01 |
| Future Conversation | 0.39 (0.20) | 0.41 (0.14) | 0.36 (0.24) | 0.009 | 0.00 |
| Highlight Benefits and Effectiveness | 0.79 (0.32) | 0.73 (0.25) | 0.84 (0.36) | 0.003 | 1.29 |
| Involve Trusted Experts | 0.12 (0.14) | 0.13 (0.14) | 0.11 (0.15) | 0.001 | 0.10 |
| Listen Actively | 0.97 (0.31) | 0.96 (0.26) | 0.99 (0.36) | 0.13 | 0.00 |
| Long-Term Health Investment | 0.15 (0.13) | 0.13 (0.10) | 0.18 (0.14) | 0.001 | 0.10 |
| Normalize the Decision | 0.02 (0.04) | 0.02 (0.03) | 0.02 (0.05) | 0.002 | 0.18 |
| Offer Resources | 0.05 (0.10) | 0.05 (0.10) | 0.05 (0.11) | 0.079 | 0.04 |
| Opportunity for Questions | 0.61 (0.22) | 0.59 (0.14) | 0.63 (0.28) | <0.001 | 0.00 |
| Patience and Reassurance | 0.60 (0.24) | 0.64 (0.18) | 0.56 (0.29) | 0.003 | 0.01 |
| Personal Stories | 0.00 (0.02) | 0.01 (0.02) | 0.00 (0.01) | <0.001 | 0.00 |
| Personalized Approach | 0.19 (0.12) | 0.22 (0.12) | 0.16 (0.12) | <0.001 | 0.02 |
| Shared Values | 0.28 (0.20) | 0.31 (0.19) | 0.25 (0.20) | <0.001 | 0.07 |
| Stay Non-Judgmental | 0.38 (0.22) | 0.42 (0.19) | 0.35 (0.24) | <0.001 | 0.03 |
| Use Analogies | 0.02 (0.05) | 0.03 (0.06) | 0.02 (0.04) | <0.001 | 0.00 |

[1]Mean (SD)
[2]Wilcoxon rank sum test

**Table S13. Linear regression predicting post-intervention HPV vaccination likelihood using six PCA-derived persuasion components**

The six components were derived from a principal component analysis of 23 persuasion strategies and together explain 62.8% of the variance in conversation-level strategy prevalence. For bipolar components, positive values correspond to the first label and negative values correspond to the opposing label (signs are arbitrary but fixed for interpretation). Coefficient signs indicate whether outcomes are more strongly associated with one pole of a component versus the other. Robust standard errors are shown below coefficients.

| | HPV Vaccine Likelihood |
|---|---|
| | |
| Conversational Chatbot (ref: Default Chatbot) | -3.19 |
| | -2.31 |
| | |
| Pre-Intervention HPV Vaccine Likelihood | 0.80*** |
| | -0.04 |
| | |
| Open Dialogue and Support | -0.88 |
| | -0.60 |
| | |
| Health Benefit Framing vs. Resources and Trusted Experts | -2.03*** |
| | -0.75 |
| | |
| Age and Timing Emphasis | 1.02 |
| | -0.80 |
| | |
| Rational Planning vs. Empathetic Personalization | 1.12 |
| | -0.92 |
| | |
| Evidence-Based Information vs. Personalized Empathy | 1.36* |
| | -0.73 |
| | |
| Herd Immunity vs. Safety Concerns | 3.95*** |
| | -1.01 |
| | |
| Constant | 17.06*** |
| | -2.01 |
| | |
| | |
| Observations | 449 |
| R2 | 0.64 |

| Adjusted R2 | 0.64 |
|---|---|
| Residual Std. Error | 23.12 (df = 440) |
| F Statistic | 99.48*** (df = 8; 440) |
| | |
| Note: | *p<0.1; **p<0.05; ***p<0.01 |

**Table S14. Persuasion Strategy LASSO Regression Coefficients**

| Feature | Coefficient |
|---|---|
| (Intercept) | 47.11 |
| Personal Stories | 49.93 |
| Normalize the Decision | 31.45 |
| Involve Trusted Experts | 30.98 |
| Opportunity for Questions | 16.62 |
| Discuss Age Appropriateness | 15.50 |
| Focus on Cancer Prevention | 12.84 |
| Patience and Reassurance | 10.36 |
| Discuss Herd Immunity | 7.98 |
| Shared Values | 5.24 |
| Pre-Intervention HPV Vaccine Likelihood | 0.75 |
| Stay Non-Judgmental | 0.00 |
| Address Safety Concerns | -0.14 |
| Use Analogies | -0.50 |
| Personalized Approach | -3.46 |
| Conversational Chatbot (ref: Default Chatbot) | -4.69 |
| Discuss Disease Prevalence | -5.26 |
| Highlight Benefits and Effectiveness | -5.58 |
| Long-Term Health Investment | -5.88 |
| Address Timing Concerns | -6.97 |
| Listen Actively | -10.92 |
| Evidence-Based Information | -13.96 |
| Offer Resources | -14.36 |
| Correct Misconceptions | -16.09 |
| Future Conversation | -18.36 |
| Empathy and Understanding | -23.54 |

**Table S15. Motivational Interviewing Technique Usage by Chatbot Variant**
Usage is scored on a four point scale by GPT-4o-mini at the conversation level (0: Low, 4: High)

| **Motivational Interviewing Technique** | **Overall** N = 449[1] | **Default Chatbot** N = 223[1] | **Conversational Chatbot** N = 226[1] | **p-value**[2] |
|---|---|---|---|---|
| Overall Motivational Interviewing Adherence | 1.92 (0.29) | 2.00 (0.13) | 1.83 (0.37) | <0.001 |
| Open Ended Questions | 1.89 (0.35) | 2.00 (0.21) | 1.78 (0.42) | <0.001 |
| Reflective Listening | 1.77 (0.51) | 1.86 (0.41) | 1.68 (0.57) | <0.001 |
| Affirmation | 1.73 (0.62) | 1.99 (0.51) | 1.48 (0.61) | <0.001 |
| Sharing Knowledge | 2.62 (0.49) | 2.74 (0.44) | 2.51 (0.51) | <0.001 |
| Summarizing | 0.95 (0.53) | 1.15 (0.43) | 0.75 (0.54) | <0.001 |

[1]Mean (SD)
[2]Wilcoxon rank sum test

**Table S16. Motivational Interviewing Technique Regression Results**
Robust standard errors below coefficients.

| | HPV Vaccine Likelihood |
|---|---|
| | |
| Conversational Chatbot (Ref: Default Chatbot) | 2.08 |
| | -2.71 |
| | |
| Pre-Intervention HPV Vaccine Likelihood | 0.81*** |
| | -0.04 |
| | |
| Affirmation | 10.18*** |
| | -2.36 |
| | |
| Open Ended Questions | -2.79 |
| | -4.60 |
| | |
| Overall MI Adherence | -2.53 |
| | -5.66 |
| | |
| Reflective Listening | -4.08 |
| | -2.84 |
| | |
| Sharing Knowledge | 7.83*** |
| | -2.42 |
| | |
| Summarizing | 0.21 |
| | -2.74 |
| | |
| Constant | -6.72 |
| | -8.43 |
| | |
| | |
| Observations | 449 |
| R2 | 0.64 |
| Adjusted R2 | 0.64 |
| Residual Std. Error | 23.21 (df = 440) |
| F Statistic | 98.30*** (df = 8; 440) |
| | |
| Note: | *p<0.1; **p<0.05; ***p<0.01 |